\documentclass[onecolumn,secnumarabic,nobibnotes,superscriptaddress,prb]{revtex4-2}

\usepackage{color, xcolor, colortbl}
\usepackage{graphicx}
\usepackage{gensymb}
\usepackage{geometry}
\usepackage{amsthm,amsmath,amssymb,amsfonts}
\usepackage{dcolumn}
\usepackage{url}
\usepackage{epstopdf}
\usepackage{enumitem}
\usepackage{algorithm}
\usepackage{algpseudocode}
\usepackage{bm}
\usepackage{appendix}
\usepackage{multirow}
\usepackage{braket}
\usepackage[english]{babel}
\usepackage{hyperref}
\usepackage[capitalize]{cleveref}
\usepackage{xpatch}
\usepackage{tikz}
\usepackage{adjustbox}
\usepackage{xspace}
\usetikzlibrary{quantikz2}
\newcommand{\bvec}[1]{\mathbf{#1}}

\newcommand{\vb}{\bvec{b}}

\newcommand{\vk}{\bvec{k}}

\newcommand{\vq}{\bvec{q}}
\newcommand{\vrr}{\bvec{r}}

\newcommand{\vG}{\bvec{G}}

\newcommand{\vK}{\bvec{K}}

\newcommand{\conj}[1]{\overline{#1}}

\newcommand{\Tr}{\operatorname{Tr}}

\usepackage{placeins}

\usepackage{xcolor}
\definecolor{purp}{RGB}{160, 32, 240}

\usetikzlibrary{fit}
\tikzset{%
  highlight/.style={rectangle,rounded corners,fill=blue!15,draw,fill opacity=0.3,thick,inner sep=0pt}
}

\global\long\def\Tr{\mathrm{Tr}}

\usepackage[percent]{overpic} 

\usepackage{listings}
\usepackage{xcolor}

\usepackage{pifont}
\usepackage{siunitx}

\definecolor{codegreen}{rgb}{0,0.6,0}
\definecolor{codegray}{rgb}{0.5,0.5,0.5}
\definecolor{codepurple}{rgb}{0.58,0,0.82}
\definecolor{backcolour}{rgb}{0.95,0.95,0.92}

\lstdefinestyle{mystyle}{
  backgroundcolor=\color{backcolour},   
  commentstyle=\color{codegreen},
  keywordstyle=\color{magenta},
  numberstyle=\tiny\color{codegray},
  stringstyle=\color{codepurple},
  basicstyle=\linespread{0.8}\ttfamily\footnotesize,
  breakatwhitespace=false,         
  breaklines=true,                 
  captionpos=b,                    
  keepspaces=true,                 
  numbers=left,                    
  numbersep=5pt,                  
  showspaces=false,                
  showstringspaces=false,
  showtabs=false,                  
  tabsize=2
}

\usetikzlibrary{shapes.geometric, arrows}

\usepackage{svg}
\usepackage{caption}
\usepackage{subcaption}

\usepackage{xr}

\newcommand{\moire}{\ensuremath{\mathrm{moir\acute{e}}}}

\begin{document}

\newcommand{\DeptMath}{Department of Mathematics, University of California, Berkeley, California 94720 USA}
\newcommand{\DeptPhys}{Department of Physics, University of California, Berkeley, California 94720 USA}
\newcommand{\LBLMath}{Applied Mathematics and Computational Research Division, Lawrence Berkeley National Laboratory, Berkeley, CA 94720, USA}
\newcommand{\LBLMater}{Materials Sciences Division, Lawrence Berkeley National Laboratory, Berkeley, CA 94720, USA}

\title{Ab initio quantum embedding description of magic angle twisted bilayer graphene at even-integer fillings}
\author{Raehyun Kim}
\thanks{These authors contributed equally.}
\affiliation{\DeptMath}
\author{Woochang Kim}
\thanks{These authors contributed equally.}
\affiliation{\DeptPhys}
\affiliation{\LBLMater}
\author{Kevin D. Stubbs}
\thanks{These authors contributed equally.}
\affiliation{\DeptMath}
\author{Steven G. Louie}
\affiliation{\DeptPhys}
\affiliation{\LBLMater}
\author{Lin Lin}
\email{linlin@math.berkeley.edu}
\affiliation{\DeptMath}
\affiliation{\LBLMath}
\date{\today}

\begin{abstract}
  Magic-angle twisted bilayer graphene (MATBG) hosts narrow moir{\'e} bands with meV-scale energy splittings, making its correlated phases sensitive to both material parameters and modeling choices in low-energy downfolding.
  We develop an \textit{ab initio} quantum-embedding workflow that derives interacting flat-band Hamiltonians from Kohn-Sham density functional theory (KS-DFT) of a relaxed, unstrained structure. Our model combines constrained random phase approximation (cRPA) screening, controlled double-counting subtraction, and an automated gauge-fixing procedure based on the selected columns of the density matrix (SCDM) that is compatible with symmetry-resolved many-body calculations. 
  Solving the resulting models using Hartree-Fock (HF) and coupled cluster singles and doubles (CCSD), we recover robust insulating Kramers intervalley coherent (KIVC) states at charge neutrality ($\nu=0$) and at electron doping ($\nu=+2$). The main new physical effect appears on the hole-doped side: at $\nu=-2$ we observe a fragile semimetal with a weak $\sqrt{3}\times\sqrt{3}$ Kekul\'e modulation and enhanced intervalley-scattering peaks in the Fourier-transformed local density of states.
  Although the underlying KS-DFT band structure is nearly particle-hole symmetric, the effective interacting Hamiltonian exhibits a pronounced particle-hole asymmetry at $\nu=\pm 2$ that we trace to momentum-dependent single-particle renormalizations generated by subtraction terms constructed from reference densities consistent with the KS-DFT filling.
  Our work provides a first-principles route for connecting microscopic electronic structure, screened interactions, subtraction choices, and scanning tunneling microscopy signatures in MATBG.
\end{abstract}

\maketitle

\section{Introduction}
\label{sec:introduction}

Moir{\'e} materials have emerged as a versatile setting to study correlation-driven phenomena, including correlated insulators, superconductivity, Chern and fractional Chern phases, and symmetry-breaking electronic textures~\cite{xie_fractional_2021,Nuckolls2020_Chern_B,nuckolls2023Quantumtextures,ColumbiaSTM,PrincetonSTM,CaltechSTM,cao_correlated_2018,PabloSC,PabloNematic,Yankowitz2019,sharpe2019emergent,YoungQAH,efetov2019,Efetov2020Screening,wu2021chern,wang2022strain,kazmierczak2021strainfield,pierce_unconventional_2021,stepanov_competing_2021,jaoui_quantum_2022,saito_independent_2020,saito_hofstadter_2021,oh_evidence_2021,wong_cascade_2020,park_flavour_2021,zondiner_cascade_2020,yu_correlated_2022,choi2021_STM}.
A defining feature of moir{\'e} materials is the appearance of a long-wavelength ``moir{\'e} superlattice'' formed by lattice mismatch, for example, through twisting or by combining distinct two-dimensional (2D) crystals.
The resulting moir{\'e} length scale can exceed the atomic scale by one to two orders of magnitude, leading to an enlarged unit cell and narrow, topologically nontrivial bands near the Fermi level.
An especially powerful experimental control is the ability to tune the filling factor continuously across wide intervals of carrier density, revealing a rich sequence of phases and transitions~\cite{PabloSC,cao_correlated_2018,Yankowitz2019,serlin2020intrinsic,tang2020simulation,regan2020mott}.

The current theoretical picture has been shaped in part by analogies to other strongly correlated materials, including the high-$T_c$ cuprates.
In that setting, understanding moir{\'e} platforms can be separated into two intertwined tasks.
The first is to qualitatively identify the low-energy competing phases, and to map out how the competition depends on a small set of materials-dependent parameters.
The second is to derive effective interacting Hamiltonians whose single-particle structure and screened Coulomb interactions are quantitatively faithful to a given device, so that the energetic ordering among nearby phases can be assessed with minimal ambiguity.
The tension between these tasks is particularly sharp in moir{\'e} systems because they combine multiple length scales with strong sensitivity to details such as screening environment, relaxation, and symmetry-breaking fields.

Direct many-body calculations at the moir{\'e} scale are infeasible, and realistic modeling therefore requires a reduced-order description that retains only the relevant low-energy electronic structure and the screened interactions among the corresponding degrees of freedom.
A widely used workflow constructs such models in two steps.
First, one derives an effective single-particle ``continuum'' Hamiltonian that captures the moir{\'e}-scale band structure while smoothing atomic-scale details~\cite{bistritzer2011moire,carr2019exact,kong2025interacting,kang2018symmetry}.
Second, one projects electron-electron interactions onto a set of narrow bands near the Fermi level, yielding an effective interacting Hamiltonian that can be studied using Hartree-Fock, density matrix renormalization group, exact diagonalization, or other many-body techniques.
This approach has been successful in organizing qualitative phenomena, and in twisted bilayer graphene it has led to concrete predictions for candidate insulating and symmetry-breaking phases, including intervalley-coherent states and charge-ordered textures~\cite{bultinck2020ground,kang2019StrongCouplingPhases,lian2020tbg,xie2020NatureCorrelatedInsulator,kwan2021KekulSpiralOrder}.
However, quantitative predictions can be sensitive to modeling choices made at each step: the construction of the continuum description, the definition of the low-energy basis, the treatment of screening, and the handling of double counting when combining band-structure information with explicit interactions.

In this work, we develop a first-principles downfolding strategy designed to reduce such sources of discrepancy while preserving the conceptual clarity of low-energy modeling.
Our starting point is Kohn-Sham density functional theory (KS-DFT), from which we extract an \textit{ab initio} description of the moir{\'e}-scale band structure for a relaxed, unstrained structure.
We then construct an interacting model by projecting onto the active manifold and embedding its interactions in an \textit{ab initio} environment.
This projection-and-screening construction can be viewed as a quantum embedding in which the active flat-band subspace is treated explicitly, while the remaining electronic degrees of freedom enter through screened interactions and subtraction terms.
Screened Coulomb matrix elements are obtained from the constrained random phase approximation (cRPA)~\cite{Aryasetiawan2004crpa}, and the resulting Hamiltonian is corrected by a double-counting subtraction whose reference density turns out to be quantitatively and qualitatively important.
Because many competing orders in moir{\'e} materials are diagnosed through their symmetry and spatial structure, an uncontrolled Bloch-orbital gauge can obstruct both initialization and interpretation of symmetry-broken mean-field solutions.
We therefore introduce an automated gauge-fixing procedure based on the selected columns of the density matrix (SCDM) method~\cite{damle2015compressed,DamleLinYing2017,DamleLin2018}, yielding a low-energy basis aligned with the relevant valley and sublattice degrees of freedom.

We apply the resulting framework to MATBG and study the charge-neutral state and the doped even-integer fillings $\nu=\pm 2$ using Hartree-Fock (HF) and coupled cluster (CCSD) theories.
At $\nu=0$ we find an insulating Kramers intervalley coherent (KIVC) solution with small post-HF correlation energy. This is consistent with previous numerical studies~\cite{bultinck2020ground,FaulstichStubbsZhuEtAl2023,xiao2025correlation} as well as theoretical characterizations of the ground state manifold in an idealized flatband interacting Hamiltonian model~\cite{kang2019strong,bultinck2020ground,lian2021TwistedBilayerGraphene,becker2025exact,StubbsBeckerLin2025,stubbs2025many}.
At $\nu=+2$ we likewise obtain an insulating KIVC solution.
These $\nu=0$ and $\nu=+2$ results therefore serve mainly as benchmarks for the embedding workflow against the established KIVC phenomenology of recent continuum and downfolded calculations. The main new physical output of the present calculation is instead on the hole-doped side: at $\nu=-2$ we find a fragile semimetallic solution, even though the underlying KS-DFT band structure is nearly particle-hole symmetric, in contrast to the insulating behavior typically obtained in recent continuum and downfolded descriptions of the unstrained problem~\cite{bultinck2020ground,khalaf2020soft}.
We trace this particle-hole asymmetry to momentum-dependent single-particle renormalizations introduced by the subtraction Hamiltonian when it is constructed from an active-space reference density $P_{\mathrm{ref}}$ consistent with the KS-DFT filling.
In particular, the subtraction produces a pronounced upward shift of the valence manifold near the moir{\'e} $\gamma$ point, strongly reducing the relevant indirect gap on the hole-doped side.
The resulting $\nu=-2$ semimetal is accompanied by a weak $\sqrt{3}\times\sqrt{3}$ Kekul\'e modulation and enhanced intervalley-scattering peaks in the FT-LDOS, consistent with STM experiments~\cite{nuckolls2023Quantumtextures}.
Our findings do not imply that $\nu=-2$ must be metallic, especially given that there have been very few ultra-low-strain devices studied experimentally.
Rather, they suggest that extrinsic effects such as heterostrain can be decisive in stabilizing a gapped state in experimental devices, and that subtraction choices can qualitatively affect the hole-side spectrum at meV energy scales.

\section{Ab Initio Quantum Embedding Method}\label{sec:method}

To characterize the electronic properties of twisted bilayer graphene (TBG), we define three types of periodic cells. First, we consider the primitive unit cells of the individual top and bottom graphene layers, each containing two carbon atoms. In TBG, the orientations of these unit cells and their corresponding Brillouin zones (BZs) are rotated by the twist angle $\theta$, as illustrated in \cref{fig:DFT}(b). At specific twist angles, the lattice vectors of the two layers satisfy a commensurability condition, forming a commensurate moir{\'e} superlattice (MSL). Finally, to model the bulk system, we define a Born--von Karman (BvK) supercell containing $N_{\bvec{k}}$ moir{\'e} unit cells. By exploiting the translational symmetry of the MSL, we simulate this BvK supercell via a sampling of $N_{\bvec{k}}$ $\vk$-points within the moir{\'e} BZ.

Performing first-principles DFT calculations for this geometry yields the Kohn-Sham (KS) eigenvalues and orbitals $\big(\epsilon_{m \vk}, \psi_{m \vk}\big)$, where $m$ is the band index and $\vk$ is the crystal momentum in the moir{\'e} BZ. We write the Bloch functions as $\psi_{m\vk}(\vrr)=e^{i\vk\cdot\vrr}u_{m\vk}(\vrr)$ with $u_{m\vk}$ periodic on the moir{\'e} unit cell $\Omega$. These KS orbitals preserve atomic-scale details of the system, which distinguishes our approach from continuum-model descriptions and serves as the starting point for constructing the downfolded interacting model.

To specify the embedding construction, we define the interacting Hamiltonian in \cref{sec:interacting-hamiltonian} and the associated double-counting subtraction in \cref{sec:double-count-subtr}.
We then introduce an automated gauge-fixing procedure based on the selected columns of the density matrix (SCDM) in \cref{sec:adapt-gauge-fixing}.
Additional details are collected in \cref{sec:dft-calc-details}.

\subsection{Construction of the Downfolded Interacting Hamiltonian}
\label{sec:interacting-hamiltonian}
For each orbital \(\psi_{m \bvec{k}}\), let \(\hat{f}_{m \bvec{k}}^{\dagger}\) and \(\hat{f}_{m \bvec{k}}\) denote the creation and annihilation operators for orbital \(\psi_{m \bvec{k}}\) respectively.
The operators \(\hat{f}_{m \bvec{k}}^{\dagger}\) and \(\hat{f}_{m \bvec{k}}\) satisfy the canonical anticommutation relations as well as the periodic boundary conditions \(\hat{f}_{m (\bvec{k} + \bvec{G})}^{\dagger} = \hat{f}_{m \bvec{k}}^{\dagger}\) and \(\hat{f}_{m (\bvec{k} + \bvec{G})} = \hat{f}_{m \bvec{k}}\) for any reciprocal lattice vector \(\bvec{G}\) of the moir{\'e} superlattice.
We define the interacting moir{\'e} Hamiltonian as:
\begin{equation}
  \hat{H}_{\moire} := \hat{H}_{0} + \hat{H}_{I} - \hat{H}_{\mathrm{sub}}.
\end{equation}
Here \(\hat{H}_{0}\) denotes the DFT single-particle band structure and \(\hat{H}_{I}\) denotes the electron-electron interactions projected onto the DFT bands.
The final term \(\hat{H}_{\mathrm{sub}}\) denotes the double-counting subtraction Hamiltonian.
Since DFT captures at least part of the electron-electron interactions, the sum of operators \(\hat{H}_{0} + \hat{H}_{I}\) includes the contributions from the electron-electron interactions twice (once in the DFT bands and once in \(\hat{H}_{I}\)).
Therefore, to obtain physically realistic results, we must subtract these double-counted interactions.
For the moment, let \(\hat{H}_{\mathrm{sub}}\) be an arbitrary \(\bvec{k}\)-resolved single-particle Hamiltonian; we return to its construction in \cref{sec:double-count-subtr}.

In terms of the creation and annihilation operators, \(\hat{H}_{0}\) and \(\hat{H}_{\mathrm{sub}}\) can be written as
\begin{equation}
  \hat{H}_{0}   = \sum_{\bvec{k}, m}^{} \epsilon_{m \bvec{k}} \hat{f}_{m \bvec{k}}^{\dagger} \hat{f}_{m \bvec{k}}  \quad \quad
  \hat{H}_{\mathrm{sub}}  = \sum_{\bvec{k}, m, n}^{} [H_{\mathrm{sub}}(\bvec{k})]_{mn} \hat{f}_{m \bvec{k}}^{\dagger} \hat{f}_{n \bvec{k}}
\end{equation}

The electron-electron interactions projected onto the DFT bands can be written in terms of the screened Coulomb kernel and overlaps of the KS orbitals.
More specifically, we define the reciprocal-space pair product \(\varrho_{m \bvec{k}, n \bvec{k}'}(\bvec{G})\):
\begin{equation}
  \label{eq:pair-product-def}
  \varrho_{m \bvec{k}, n \bvec{k}'}(\bvec{G}) := \int_{\Omega}^{} e^{-i \bvec{G} \cdot \bvec{r}} e^{i (\bvec{k} - \bvec{k}') \cdot \bvec{r}}  \conj{\psi_{m \bvec{k}}(\bvec{r})} \psi_{n \bvec{k}'}(\bvec{r}) \mathrm{d}\bvec{r}.
\end{equation}
With this definition, \(\hat{H}_{I}\) can be written
\begin{equation}
  \label{eq:h-interacting}
  \begin{split}
    \hat{H}_{I}
    & = \frac{1}{2 N_{\bvec{k}} |\Omega|} \sum_{\bvec{k}, \bvec{k}', \bvec{q}}^{} \sum_{\bvec{G}, \bvec{G}'}^{} \sum_{mnm'n'}^{} \bigg\{  W(\bvec{q}; \bvec{G}, \bvec{G}') \\
    & \hspace{8em}  \varrho_{m \bvec{k}, n (\bvec{k} + \bvec{q})}(\bvec{G}) \varrho_{m' \bvec{k}', n' (\bvec{k}'- \bvec{q})}(-\bvec{G}') \hat{f}_{m \bvec{k}}^{\dagger} \hat{f}_{m' \bvec{k}'}^{\dagger} \hat{f}_{n' (\bvec{k}' - \bvec{q})} \hat{f}_{n (\bvec{k} + \bvec{q})}  \bigg\}.
  \end{split}
\end{equation}
where \(W(\bvec{q}; \bvec{G}, \bvec{G}')\) denotes the screened Coulomb kernel.
In this work, we compute $W$ using the (static) constrained random phase approximation (cRPA), in which screening processes internal to the chosen active (flat-band) manifold are excluded to avoid double counting.
Denoting by $\chi_0(\bvec{q};\bvec{G},\bvec{G}')$ the full irreducible polarizability and by $\chi_A(\bvec{q};\bvec{G},\bvec{G}')$ the contribution from virtual transitions within the active manifold, we define the rest polarizability
\begin{equation}
  \chi_R(\bvec{q};\bvec{G},\bvec{G}') := \chi_0(\bvec{q};\bvec{G},\bvec{G}') - \chi_A(\bvec{q};\bvec{G},\bvec{G}').
\end{equation}
The screened interaction entering \cref{eq:h-interacting} is then written in terms of the cRPA dielectric matrix as
\begin{equation}
  W(\bvec{q};\bvec{G},\bvec{G}') := \varepsilon_{\rm cRPA}^{-1}(\bvec{q};\bvec{G},\bvec{G}')\, V_{\mathrm{slab}}(\bvec{q}+\bvec{G}').
\end{equation}
Here $V_{\mathrm{slab}}$ denotes the truncated Coulomb kernel appropriate to the slab geometry. Further details of the cRPA construction and numerical approximations are given in \cref{sec:scre-coul-inter}.

In the moir{\'e} literature, the expression for \(\hat{H}_{I}\) is often written in terms of the form factor \(\Lambda_{\bvec{k}}(\bvec{q} + \bvec{G})\).
In this work, we use the pair product to match more closely with the quantum chemistry literature.
The form factor is a notational variant of the pair product defined in \cref{eq:pair-product-def}, and one can verify that \cref{eq:h-interacting} is equivalent to standard projected-interaction formulas~\cite{lian2020tbg,bultinck2020ground,ledwith2021strong}.

\subsection{Double-Counting Subtraction}
\label{sec:double-count-subtr}
As discussed above, for defining an interacting Hamiltonian based on \textit{ab initio} DFT calculations, it is important to remove the electron-electron interaction effects that are already implicitly included in the DFT band energies through the Hartree and exchange-correlation (XC) potentials.
A common choice for double-counting subtraction in the interacting moir{\'e} literature is to use a mean-field Hamiltonian of the following form~\cite{Kang2020_C2T,bernevig2021twisted,bultinck2020ground,Soejima2020_efficient,xie2020NatureCorrelatedInsulator,FaulstichStubbsZhuEtAl2023}:
\begin{equation}
  \label{eq:sub_ham_def}
  H_{\text{sub}}(\vk) := J[P_{\text{ref}}](\vk) + K[P_{\text{ref}}](\vk)
\end{equation}
where $J[P_{\mathrm{ref}}](\vk)$ and $K[P_{\mathrm{ref}}](\vk)$ denote the Coulomb and exchange contributions constructed from a reference one-body density $P_{\mathrm{ref}}$.
Exact expressions for these matrices can be derived from~\cref{eq:h-interacting} using Wick's theorem; see \cref{subsec:form-coul-exch}.

Typical choices for the reference density include the average subtraction~\cite{Kang2020_C2T,bernevig2021twisted}, charge-neutral subtraction~\cite{hejazi2021HybridWannierChern}, and decoupled subtraction~\cite{bultinck2020ground,Soejima2020_efficient,xie2020NatureCorrelatedInsulator} based on different physical considerations.
We also note the work of Kang and Vafek~\cite{vafek2020renormalization} which addresses the problem of double-counting subtraction using an approach inspired by the renormalization group.
In a first-principles embedding starting from a KS-DFT calculation at a specified filling, however, the reference density used in $H_{\mathrm{sub}}$ should represent the corresponding KS reference within the chosen active manifold.
For example, for a charge-neutral KS reference, the conduction flat bands are empty, so a particle-hole symmetric ``average'' choice that assigns equal weight to valence and conduction states does not correspond to the projected KS density in the active space.
As we show below for $\nu=\pm 2$, this distinction can generate qualitatively different momentum-dependent single-particle renormalizations in the effective Hamiltonian.

This mean-field-based subtraction may overestimate the double-counting correction associated with the KS exchange-correlation potential, since a semi-local XC functional does not reproduce the nonlocal, long-range structure of Fock exchange.
To mitigate this mismatch, we isolate the flat-band contribution to the exchange-correlation potential as follows:
\begin{align*}
  v^{r}_{xc} := v_{xc}[\rho] - v_{xc}[\rho-\rho^{f}],
\end{align*}
where $\rho$ represents the total charge density of all occupied bands, while $\rho^{f}$ denotes the density contribution from the occupied flat bands.
In this construction, $v_{xc}^{r}$ represents the differential of the XC potential induced by the flat-band density.
In the linearized limit, this term is approximated by:
\begin{align*}
  v^{r}_{xc}(\vrr) \approx \int d\vrr^{\prime} K_{xc}[\rho](\vrr,\vrr^{\prime})\rho^{f}(\vrr^{\prime}),
\end{align*}
where $K_{xc}$ is the exchange-correlation kernel, defined as the second-order functional derivative of the exchange-correlation energy with respect to the density.

Combining this with the Hartree potential computed using the screened interaction $W$, we define the DFT subtraction term $H_{\text{sub}}^{\text{DFT}}$ as:
\begin{align}
  H_{\text{sub}}^{\text{DFT}}(\vk) :=  J[P_{\text{ref}}](\vk)+v^{r}_{xc}(\vk)
\end{align}
which we refer to as the DFT subtraction scheme.
Here, $P_{\text{ref}}$ is chosen to be the density matrix of the flat-band subspace at the charge neutrality point.
The matrix elements are computed as
\begin{align*}
  \left[J[P_{\text{ref}}](\vk)\right]_{mn} &=  \sum_{\vk'} \sum_{m'n'} \braket{m \bvec{k}, m' \bvec{k}' | n \bvec{k}, n' \bvec{k}' } \left[P_{\text{ref}}(\vk')\right]_{m'n'}, \\
    [v^{r}_{xc}(\vk)]_{mn} & = \int_{\Omega} \conj{u_{m\vk}(\vrr)} v^{r}_{xc}(\vrr)u_{n\vk}(\vrr) d \vrr. 
\end{align*}
The DFT subtraction scheme aims to remove the mean-field contributions of the active (flat-band) charge density already encoded in the KS band energies, while avoiding the over-subtraction issues associated with a Hartree-Fock-based subtraction.

\subsection{Adaptive Gauge Fixing with SCDM}
\label{sec:adapt-gauge-fixing}

While the physics of the interacting Hamiltonian \(\hat{H}_{\moire}\) is gauge invariant, the choice of gauge for the Bloch orbitals can strongly affect numerical convergence and the interpretability of symmetry-broken Hartree-Fock solutions.
Starting from the KS-DFT Bloch orbitals, we use an adaptive SCDM-based localization procedure~\cite{damle2015compressed,DamleLinYing2017,DamleLin2018} to construct a valley- and sublattice-resolved basis (the also called the sublattice-polarized gauge), which provides a convenient representation for initializing and classifying symmetry-broken ansatzes.
Implementation details are provided in \cref{app:gauge-fixing-impl}; see also \Cref{alg:valley-sublattice-localization}.

After localization, we transform the Bloch orbitals \(\psi_{m \bvec{k}}\) to sublattice- and valley-polarized orbitals \(\psi_{(\sigma, \tau), \bvec{k}}\), where \(\sigma \in \{ A, B \}\) and \(\tau \in \{ \vK, \vK' \}\).
For MATBG, the DFT Hamiltonian \(H(\bvec{k})\) satisfies spinless time reversal, \(\mathcal{T}\), and $180^{\circ}$ in-plane rotation, $C_{2z}$.
We therefore enforce the following relations (up to $\bvec{k}$-dependent U(1) phases), where $-\tau$ denotes the opposite valley and $-\sigma$ denotes the opposite sublattice:
\begin{equation}
  \label{eq:tr-and-c2-gauge}
  \begin{split}
    \mathcal{T} \psi_{(\sigma, \tau), \bvec{k}}(\bvec{r}) & = \conj{\psi_{(\sigma, -\tau), (-\bvec{k})}(\bvec{r})}, \\
    C_{2z} \psi_{(\sigma, \tau), \bvec{k}}(\bvec{r}) & = \psi_{(-\sigma, \tau), \bvec{k}}(-\bvec{r}).
  \end{split}
\end{equation}

\begin{figure}[htbp]
    \centering
    \begin{subfigure}{0.32\textwidth}
        \centering
        \includegraphics[width=0.9\textwidth]{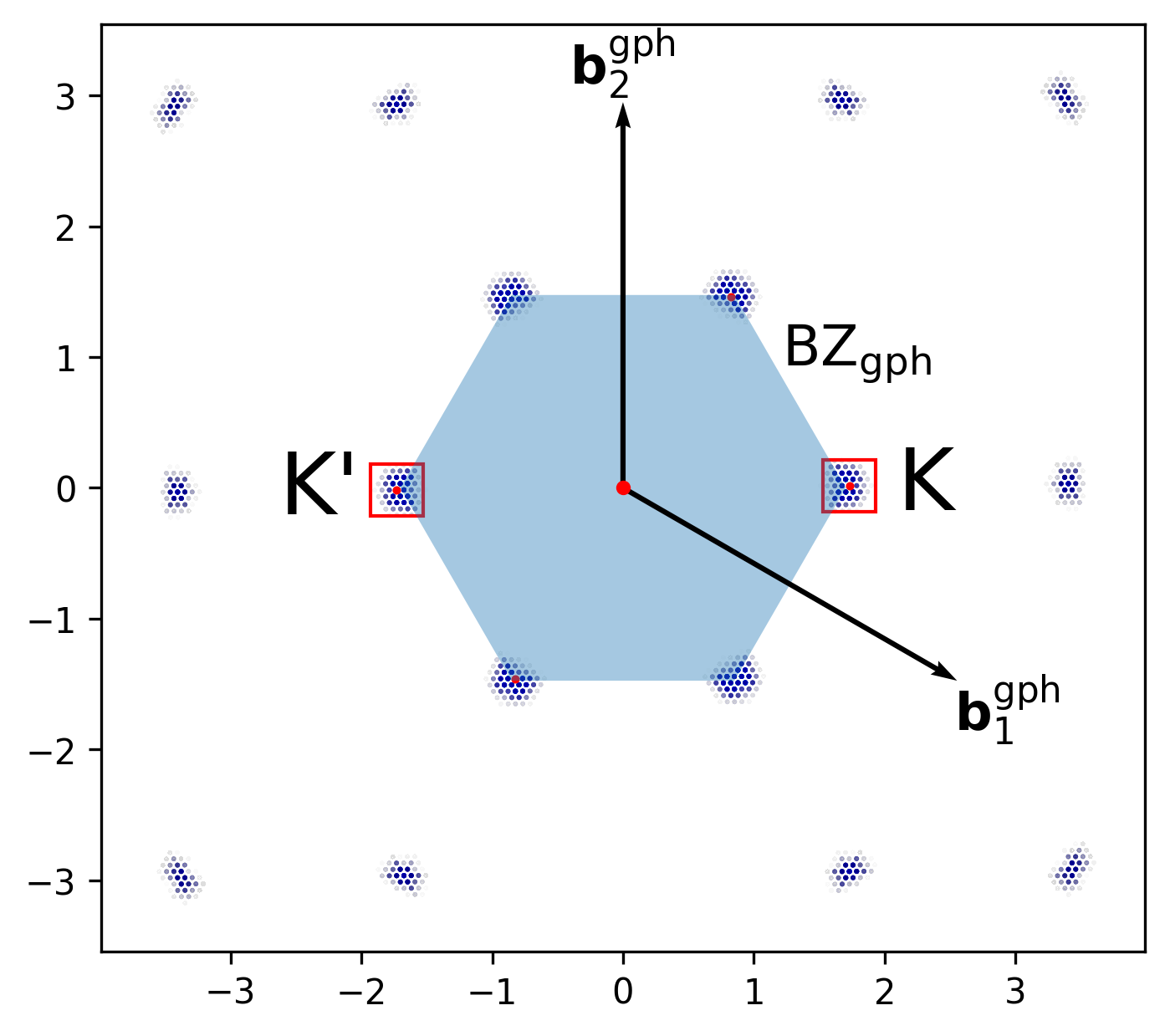}
        \caption{}
        \label{fig:val_pol_before}
    \end{subfigure}
    \begin{subfigure}{0.32\textwidth}
        \centering
        \includegraphics[width=0.9\textwidth]{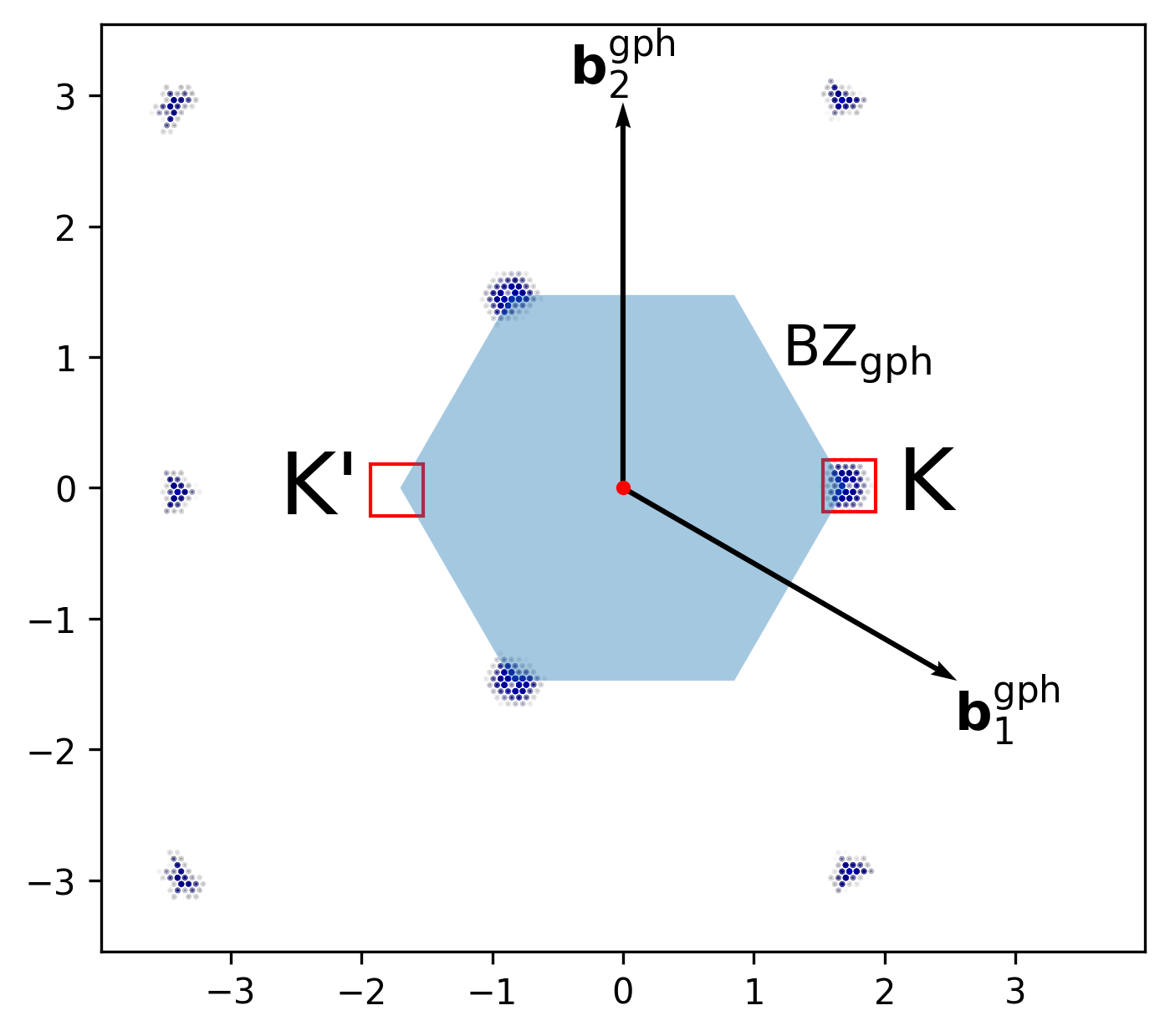}
        \caption{}
        \label{fig:val_pol_K}
    \end{subfigure}
    \begin{subfigure}{0.32\textwidth}
        \centering
        \includegraphics[width=0.9\textwidth]{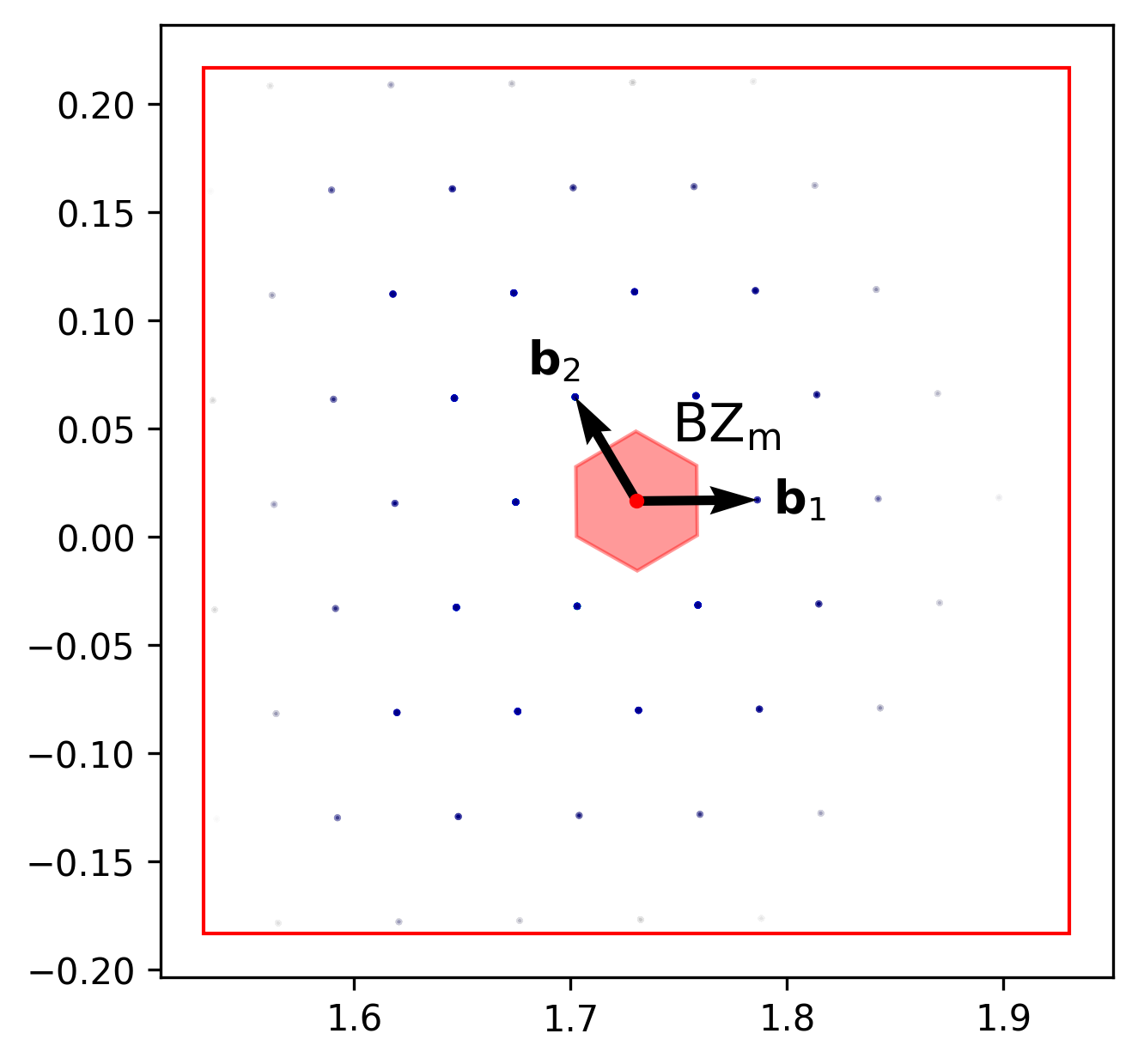}
        \caption{}
        \label{fig:val_pol_K_zoom}
    \end{subfigure}
    \caption{(\subref{fig:val_pol_before}) Superposition of the density of DFT band basis in $\vG$-space. Red dots indicate the location of four significant points from SCDM and the valley $\vK$ is specified by the highest peak in the SCDM result. Showing the graphene Brillouin zone ($\mathrm{BZ}_{\mathrm{gph}}$) and graphene lattice vectors ($\vb^{\mathrm{gph}}_1, \vb^{\mathrm{gph}}_2$) (\subref{fig:val_pol_K}) $\vK$-valley polarized basis in $\vG$-space (\subref{fig:val_pol_K_zoom}) Zoomed-in view of the $\vK$-valley center, showing the moir\'e Brillouin zone ($\mathrm{BZ}_\mathrm{m}$) and reciprocal lattice vectors ($\vb_1, \vb_2$)}
    \label{fig:val_pol}
\end{figure}

\begin{figure}[htbp]
  \centering
  \begin{subfigure}{0.32\textwidth}
    \centering
    \includegraphics[width=\linewidth]{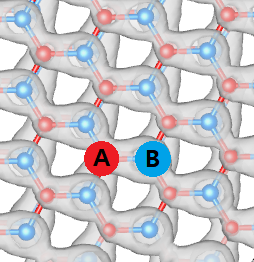}
    \caption{}
    \label{fig:sub_pol_before}
  \end{subfigure}
  \begin{subfigure}{0.32\textwidth}
    \centering
    \includegraphics[width=\linewidth]{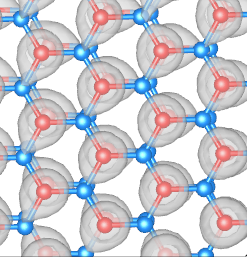}
    \caption{}
    \label{fig:sub_pol_after_A}
  \end{subfigure}
  \begin{subfigure}{0.32\textwidth}
    \centering
    \includegraphics[width=\linewidth]{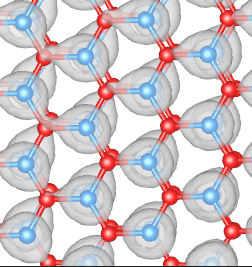}
    \caption{}
    \label{fig:sub_pol_after_B}
  \end{subfigure}
  \caption{(\subref{fig:sub_pol_before}) Superposition of the density of $\vK$-valley polarized basis in $\vrr$-space. Red and blue atoms are $A$ and $B$ sublattices, respectively. (\subref{fig:sub_pol_after_A}) $A$-sublattice-polarized basis in $\vrr$-space. (\subref{fig:sub_pol_after_B}) $B$-sublattice-polarized basis in $\vrr$-space.}
  \label{fig:sub_pol}
\end{figure}

Representative outcomes of valley and sublattice localization for MATBG are shown in \cref{fig:val_pol,fig:sub_pol}.

\section{Numerical Results}
\label{sec:results}
In this section, we report the interacting Hartree-Fock (HF) and coupled cluster singles and doubles (CCSD) results from the \textit{ab initio} quantum embedding model.
We focus on the even-integer fillings $\nu=0,\pm 2$ and diagnose candidate phases by their mean-field energetics, band structures, and scanning-tunneling-microscopy (STM) signatures.
In particular, motivated by STM observations of translation-breaking charge order~\cite{nuckolls2023Quantumtextures}, we report the local density of states (LDOS) and its Fourier transform (FT-LDOS) in addition to the HF and CCSD energies; details of the LDOS/FT-LDOS construction are given in \cref{app:dos}.

To explore the symmetry-broken landscape of this model, we employ both structured initializations (based on the standard valley polarized (VP), valley Hall (VH), quantum Hall (QH), Kramers intervalley coherent (KIVC), and time-reversal symmetric intervalley coherent (TIVC) ansatzes) 
and unbiased randomized initializations.
The explicit initialization prescriptions and symmetry diagnostics used to classify converged states are summarized in \cref{app:init}, while additional randomized-search data are collected in \cref{sec:randomized-tests}.

Among these candidates, previous continuum-model studies predict the KIVC insulating state to be the ground state at even-integer filling in the absence of strain or electron-phonon interactions~\cite{bultinck2020ground,kwan2021KekulSpiralOrder,kwan2024electron,Wagner2022}.
However, experimental observations, particularly the presence of intervalley-scattering peaks at $\nu=-2$, suggest a more complex ground-state landscape potentially involving inter-Chern intervalley coherent orders (IC-IVC)~\cite{nuckolls2023Quantumtextures}.
Motivated by these discrepancies, we compare relaxations from the standard ansatzes (\cref{eq:initial-states}) against randomized initializations to test the robustness of the KIVC solution and to identify other possible lower-energy states.

\subsection{DFT Band Structure}
\label{sec:dft-band-structure}

\begin{figure*}[ht]
    \centering
    \includegraphics[width=0.48\linewidth]{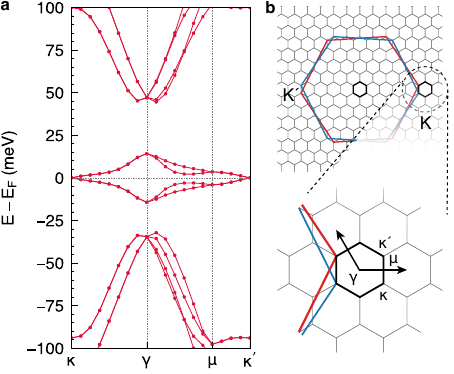}
    \caption{(a) Density functional theory (DFT) band structure of twisted bilayer graphene (TBG) at a twist angle $\theta = 1.08^{\degree}$. (b) The primitive unit-cell Brillouin zones (BZ) of the top and bottom layers of graphene with a large twist angle $\theta = 7.34^{\degree}$ are shown in red and blue, respectively, for illustrative purposes. The moir{\'e} BZ is shown in black with various high-symmetry $k$-points.}

    \label{fig:DFT}    
\end{figure*}

We begin by summarizing the KS-DFT band structure from the relaxed atomic geometry at the magic angle $\theta = 1.08^\circ$; see \cref{fig:DFT}(a).
The DFT spectrum reveals an isolated manifold of four narrow bands centered near the Fermi level, shown along the moir{\'e} high-symmetry path $\kappa$ -- $\gamma$ -- $\mu$ -- $\kappa'$.
Here lowercase Greek letters are used for moir{\'e} high-symmetry points, whereas capital Greek letters are reserved for non-moir{\'e} ones.
Due to atomic relaxation, this low-energy subspace exhibits a bandwidth of approximately $30$ meV and is clearly separated from the remote dispersive states.
The energy gap between the flat bands and remote bands is approximately $30$ meV on the electron side and $20$ meV on the hole side. Furthermore, consistent with recent DFT benchmarks~\cite{lucignano2019matbgdft,zhu2025matbgdft}, we do not observe strong particle-hole (PH) symmetry breaking in the DFT band structure, in contrast to the ``mexican hat'' feature (an inversion of band curvature away from the moir{\'e} $\gamma$ point) that appears in certain parametrizations of continuum models~\cite{kang2023pseudomagnetic,carr2019exact,QuinnKongLuskinEtAl2025}.

\subsection{Even Fillings: Robust Insulating KIVC at \texorpdfstring{$\nu=0$ and $\nu=+2$}{nu=0 and nu=+2}}
\label{sec:numerical-results-nu-0-p2}
\begin{figure*}[htbp]
    \centering
    \begin{overpic}[width=0.9\linewidth]{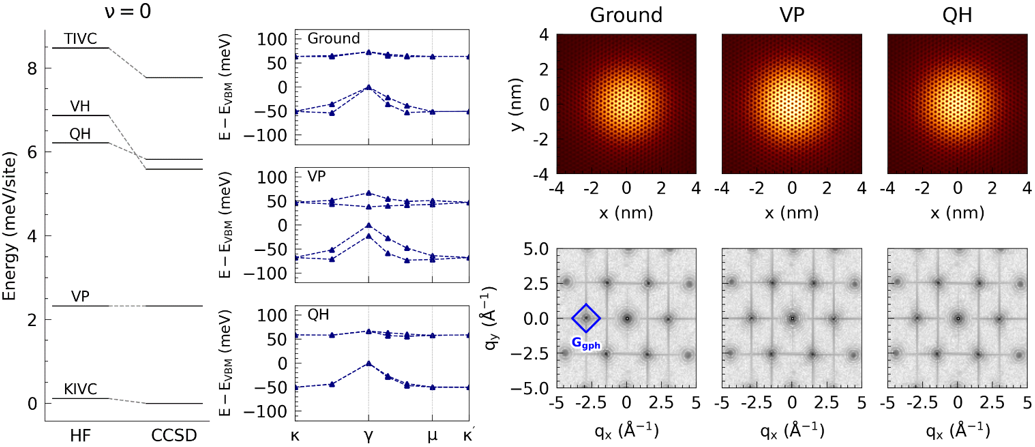}
          \put(0,40.5){\scriptsize \textcolor{black}{\textbf{a}}}  
          \put(21,40.5){\scriptsize \textcolor{black}{\textbf{b}}}  
          \put(21,27.2){\scriptsize \textcolor{black}{\textbf{c}}}  
          \put(21,14){\scriptsize \textcolor{black}{\textbf{d}}}  
          
          \put(54.5,38.4){\scriptsize \textcolor{white}{\textbf{e}}}  
          \put(70.5,38.4){\scriptsize \textcolor{white}{\textbf{f}}}  
          \put(86.5,38.4){\scriptsize \textcolor{white}{\textbf{g}}}  
          
          \put(54.5,17.4){\scriptsize \textcolor{black}{\textbf{h}}}  
          \put(70.5,17.4){\scriptsize \textcolor{black}{\textbf{i}}}  
          \put(86.5,17.4){\scriptsize \textcolor{black}{\textbf{j}}}  
    \end{overpic}
    \caption{(a) Energy hierarchy of various phases of MATBG at charge neutrality at the Hartree-Fock (HF) and coupled cluster singles and doubles (CCSD) levels. (b-d) HF band structure, (e-g) Local density of states (LDOS), (h-j) The corresponding Fourier-transformed LDOS (FT-LDOS) of the KIVC, VP, and QH states, respectively. Here, $E_{\text{VBM}}$ denotes the valence band maximum. In the FT-LDOS, the Bragg peaks of the graphene lattice are denoted by blue diamonds.}
    \label{fig:nu0}    
  \end{figure*}
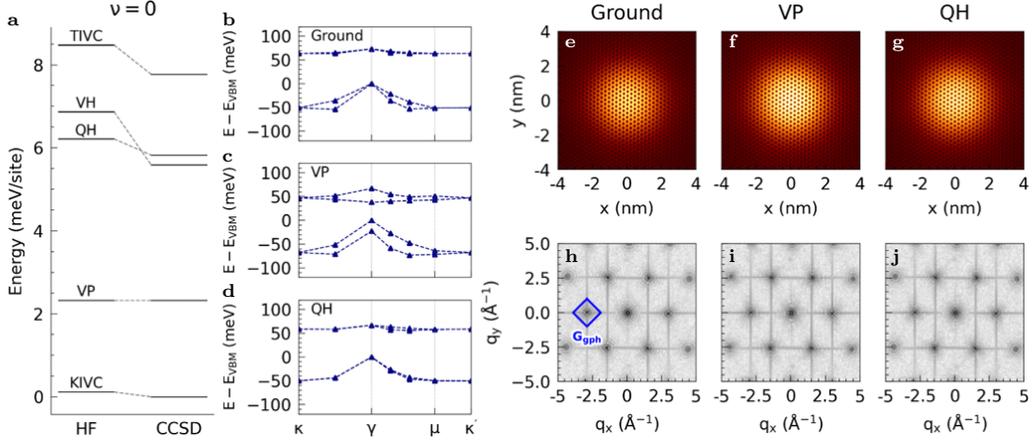

We first consider the even fillings $\nu = 0$ and $\nu = +2$. In both cases, self-consistent HF favors an insulating KIVC solution, and CCSD yields only small correlation energies beyond HF.

  \emph{Results for $\nu = 0$}: We investigate five candidate symmetry-broken solutions using self-consistent HF and CCSD calculations to determine their energy ordering and characteristics (\cref{fig:nu0}).
We find that HF yields insulating phases in contrast to the metallic or semimetallic behavior often obtained in KS-DFT~\cite{da2021correlation}, reflecting the nonlocal exchange contributions present in HF but absent in DFT and Hartree-only approaches~\cite{xie2020NatureCorrelatedInsulator,goodwin2020hartree,HouSurWagnerEtAl2025}.

The energy ordering of the HF calculations closely matches theoretical expectations based on second-order perturbation theory~\cite{bultinck2020ground}.
In particular, the KIVC state has significantly lower energy compared to the other states.
Furthermore, the CCSD correlation energies ($E_c = E_{\rm CCSD}-E_{\rm HF}$) are very small ($<\SI{0.1}{meV}$ per moir{\'e} unit cell) which is consistent with theoretical predictions at the chiral limit~\cite{stubbs2025many,bultinck2020ground,bernevig2021twisted}.

We note that the FT-LDOS of the three lowest-energy states (KIVC, VP, and QH) shows Bragg peaks associated with the graphene lattice and no additional Kekul{\'e} peaks.
This is expected since VP and QH have no intervalley component (and hence cannot break graphene translation symmetry), while for a fully gapped KIVC insulator the Kekul{\'e} signature vanishes by symmetry~\cite{HongSoejimaZaletel2022}.

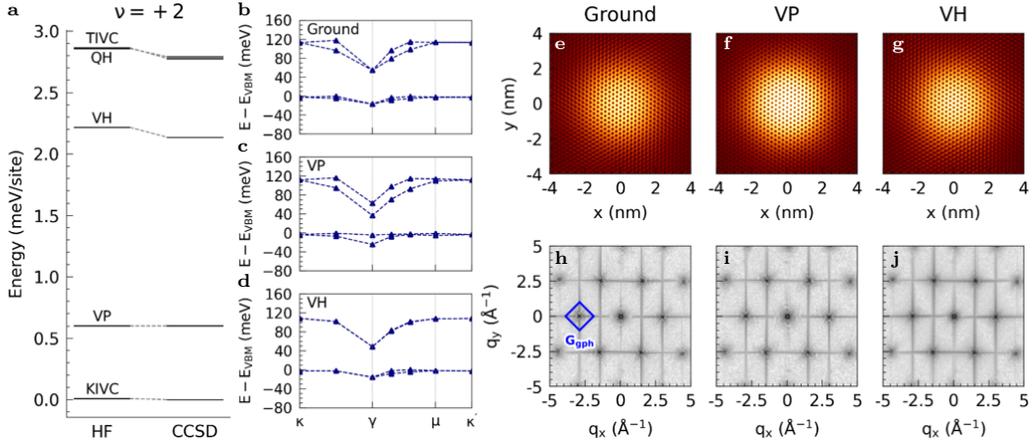
\begin{figure*}[htbp]
    \centering
    \begin{overpic}[width=0.9\linewidth]{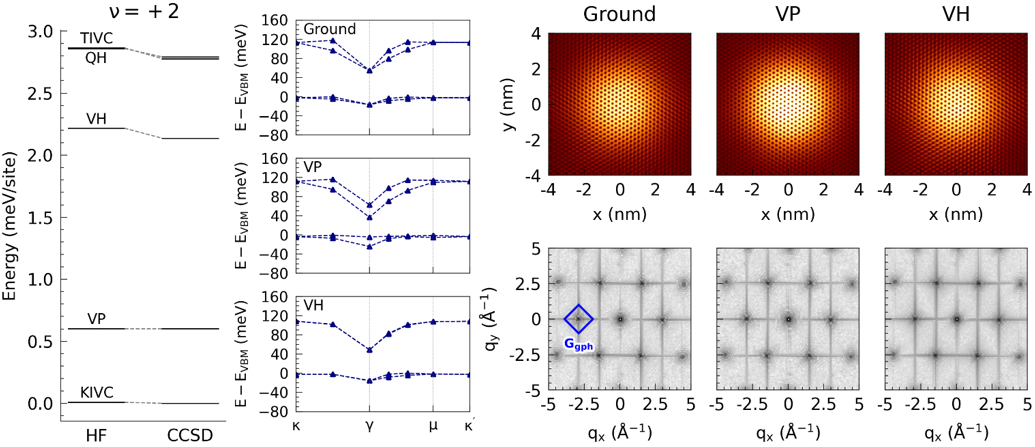}
          \put(0,41.7){\scriptsize \textcolor{black}{\textbf{a}}}  
          \put(22.5,41.7){\scriptsize \textcolor{black}{\textbf{b}}}  
          \put(22.5,28.4){\scriptsize \textcolor{black}{\textbf{c}}}  
          \put(22.5,15.2){\scriptsize \textcolor{black}{\textbf{d}}}  
          
          \put(53.5,38.4){\scriptsize \textcolor{white}{\textbf{e}}}  
          \put(70.0,38.4){\scriptsize \textcolor{white}{\textbf{f}}}  
          \put(86.5,38.4){\scriptsize \textcolor{white}{\textbf{g}}}  
          
          \put(53.5,17.4){\scriptsize \textcolor{black}{\textbf{h}}}  
          \put(70.,17.4){\scriptsize \textcolor{black}{\textbf{i}}}  
          \put(86.5,17.4){\scriptsize \textcolor{black}{\textbf{j}}}  
    \end{overpic}
    \caption{(a) Energy hierarchy of various phases at $\nu=+2$ at the HF and CCSD levels. (b-d) HF band structure, (e-g) LDOS, (h-j) The corresponding FT-LDOS of KIVC, VP, and VH state, respectively. Here, $E_{\text{VBM}}$ denotes the valence band maximum. In the FT-LDOS, the Bragg peaks of the graphene lattice is denoted with blue diamond.}
    \label{fig:nup2}    
\end{figure*}

\emph{Results for $\nu = +2$}: The corresponding HF/CCSD results are shown in \cref{fig:nup2}. The qualitative picture is unchanged from $\nu=0$: KIVC remains the lowest-energy state among the structured ansatzes, and the CCSD correction is again small. The FT-LDOS of the three lowest-energy states contains only graphene Bragg peaks, with no Kekul{\'e} features.

\subsection{Hole Doping: Fragile Semimetal and Enhanced Intervalley Peaks at \texorpdfstring{$\nu=-2$}{nu=-2}}
\label{sec:numerical-results-nu-m2}

\begin{figure*}[htbp]
    \centering
    \begin{overpic}[width=0.9\linewidth]{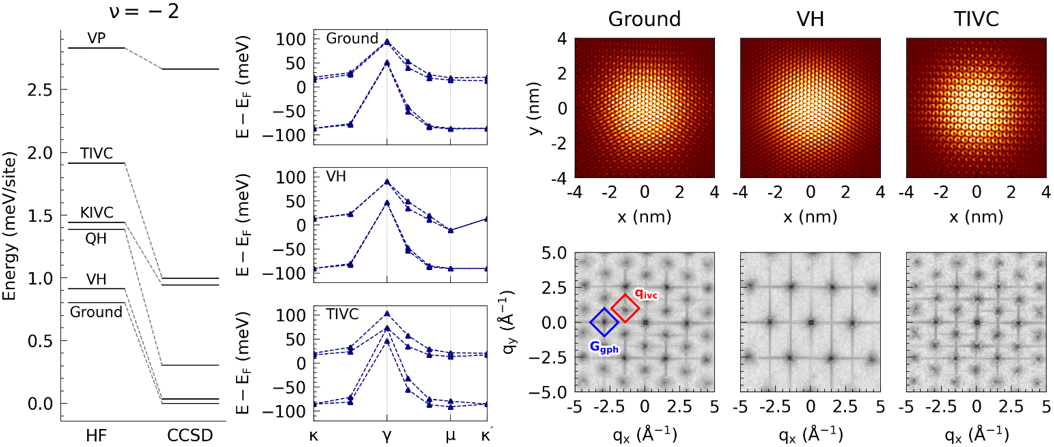}
          \put(0,40.5){\scriptsize \textcolor{black}{\textbf{a}}}  
          \put(22,40.5){\scriptsize \textcolor{black}{\textbf{b}}}  
          \put(22,27.2){\scriptsize \textcolor{black}{\textbf{c}}}  
          \put(22,14){\scriptsize \textcolor{black}{\textbf{d}}}  
          
          \put(54.7,37.2){\scriptsize \textcolor{white}{\textbf{e}}}  
          \put(70.7,37.2){\scriptsize \textcolor{white}{\textbf{f}}}  
          \put(86.7,37.2){\scriptsize \textcolor{white}{\textbf{g}}}  
          
          \put(54.7,16.8){\scriptsize \textcolor{black}{\textbf{h}}}  
          \put(70.7,16.8){\scriptsize \textcolor{black}{\textbf{i}}}  
          \put(86.7,16.8){\scriptsize \textcolor{black}{\textbf{j}}}  
    \end{overpic}
    \caption{(a) Energy hierarchy of various phases at $\nu=-2$ at the HF and CCSD levels. (b-d) HF band structure. (e-g) LDOS. (h-j) The corresponding FT-LDOS of the lowest-energy solution found in the randomized search, the TIVC state, and the VH state, respectively. Here, $E_{\text{F}}$ denotes the Fermi energy. In the FT-LDOS, the Bragg peaks of the graphene lattice and the intervalley-scattering peaks are denoted with blue and red diamonds, respectively.}
    \label{fig:nun2}    
  \end{figure*}
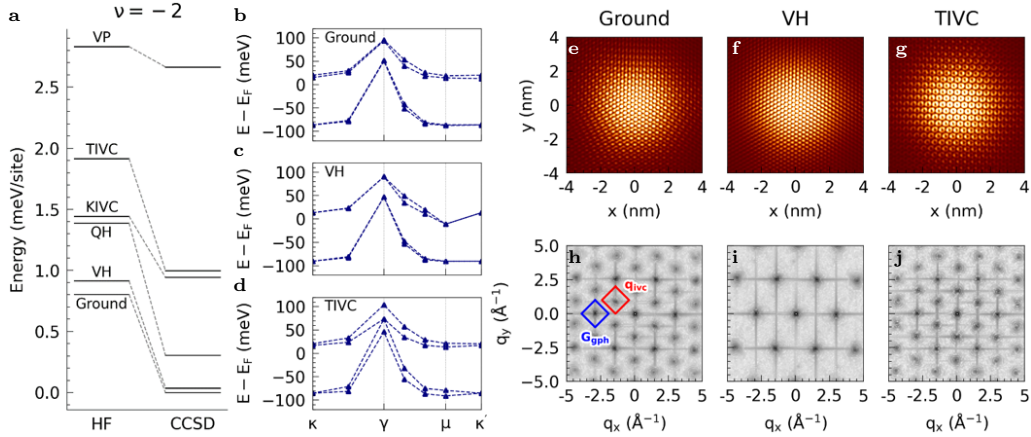

At \(\nu = -2\), our HF calculations reveal the main qualitative discrepancy between the present embedding approach and standard continuum-model expectations for unstrained MATBG~\cite{bultinck2020ground,khalaf2020soft}.
Using randomized initialization, we find a lowest-energy state distinct from the previously identified candidate states.
The energy hierarchy deviates significantly from continuum model predictions (\cref{fig:nun2} a)~\cite{bultinck2020ground,khalaf2020soft}.
We also observe that correlation energies $E_c$ increase substantially to approximately $\SI{1}{meV}$ per moir{\'e} unit cell, indicating that electronic correlations play a more important role at $\nu=-2$ compared to both \(\nu = 0\) and \(\nu = +2\).

We characterize this lowest-energy state using the same symmetry diagnostics as for the structured ansatzes (\cref{app:init}), and we refer to it as the ``ground'', i.e. lowest-energy solution found in our randomized search in \cref{fig:nun2}.

Although the energy and HF band structure of the lowest-energy state (\cref{fig:nun2} a,b) resemble those of the VH state (\cref{fig:nun2} a,d), the real-space and momentum-space characteristics are clearly distinct.
The LDOS of the lowest-energy state (\cref{fig:nun2} e) shows clear spatial modulation, in contrast to the more uniform VH state (\cref{fig:nun2} g).
The distinction is more apparent in the FT-LDOS.
While the VH state (\cref{fig:nun2} j) only shows the primary graphene Bragg peaks ($G_{\mathrm{gph}}$), the lowest-energy state (\cref{fig:nun2} h) exhibits additional, stronger intervalley-scattering peaks ($q_{\mathrm{ivc}}$), marked by red diamonds.
$q_{\mathrm{ivc}}$ represents the emergence of a $\sqrt{3} \times \sqrt{3}$ charge modulation and is consistent with the intervalley-scattering peaks reported in ultra-low-strain STM experiments~\cite{nuckolls2023Quantumtextures}.
In particular, these peaks do not by themselves uniquely determine the underlying order parameter: in a semimetallic state, the symmetry-based arguments that enforce a vanishing Kekul\'e signature for a fully gapped KIVC insulator~\cite{HongSoejimaZaletel2022} need not apply.

Regarding the electronic band structure, we find that all low-energy candidate states shown in \cref{fig:nun2}, including the lowest-energy state found via randomized initialization, are semimetallic (\cref{fig:nun2} b-d) with bands crossing the Fermi level.
This gapless behavior stands in contrast to continuum model studies, which typically predict a robust insulating state at $\nu = -2$ in the unstrained setting~\cite{kwan2021KekulSpiralOrder,bultinck2020ground}.
We find that this semimetallic behavior is largely controlled by the double-counting subtraction procedure in the effective Hamiltonian, as discussed next.

\subsection{Subtraction Induced Particle-Hole Asymmetry}
\label{sec:density-ph-asymmetry}

\begin{figure*}[h]
  \centering
  \includegraphics[width=.6\linewidth]{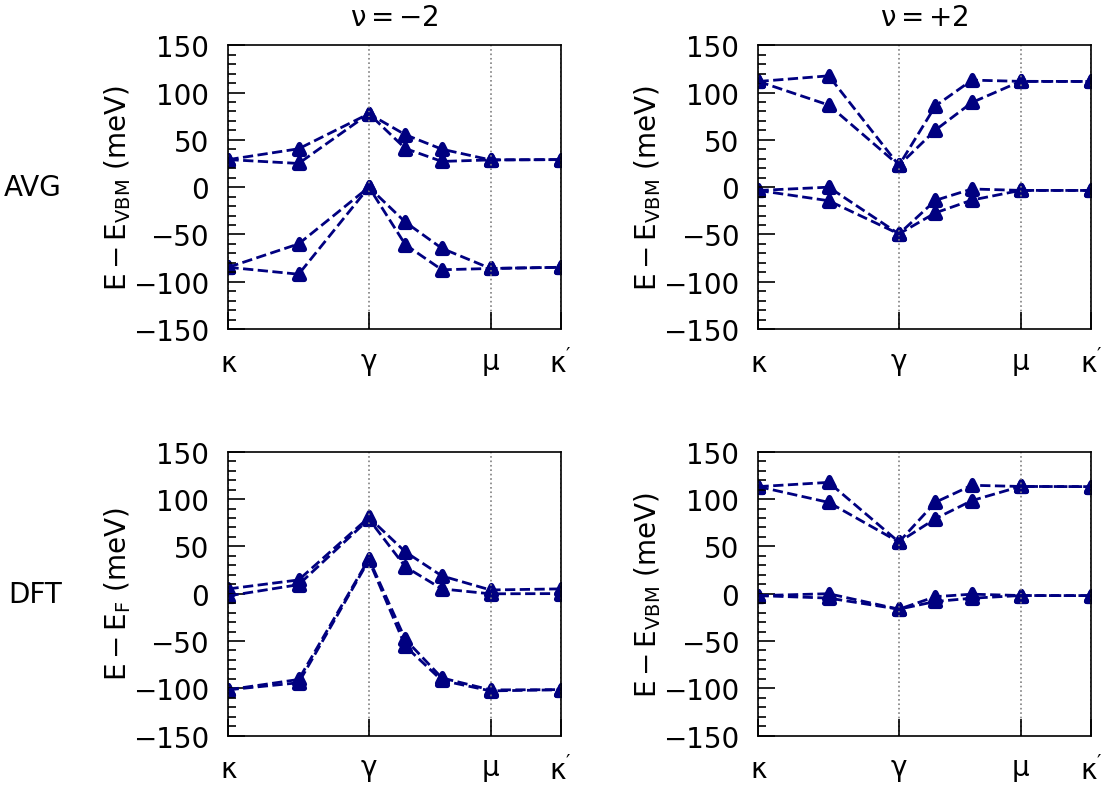}
  \caption{Comparison between HF band structures obtained with (top) a particle-hole symmetric ``average'' subtraction and (bottom) a DFT subtraction constructed from the projected charge-neutral reference density. Here, $E_{\text{VBM}}$ and $E_F$ denote the valence band maximum and the Fermi energy, respectively.}
  \label{fig:subtraction_comparison}
\end{figure*}

\begin{figure*}[h]
  \centering
  \includegraphics[width=.6\linewidth]{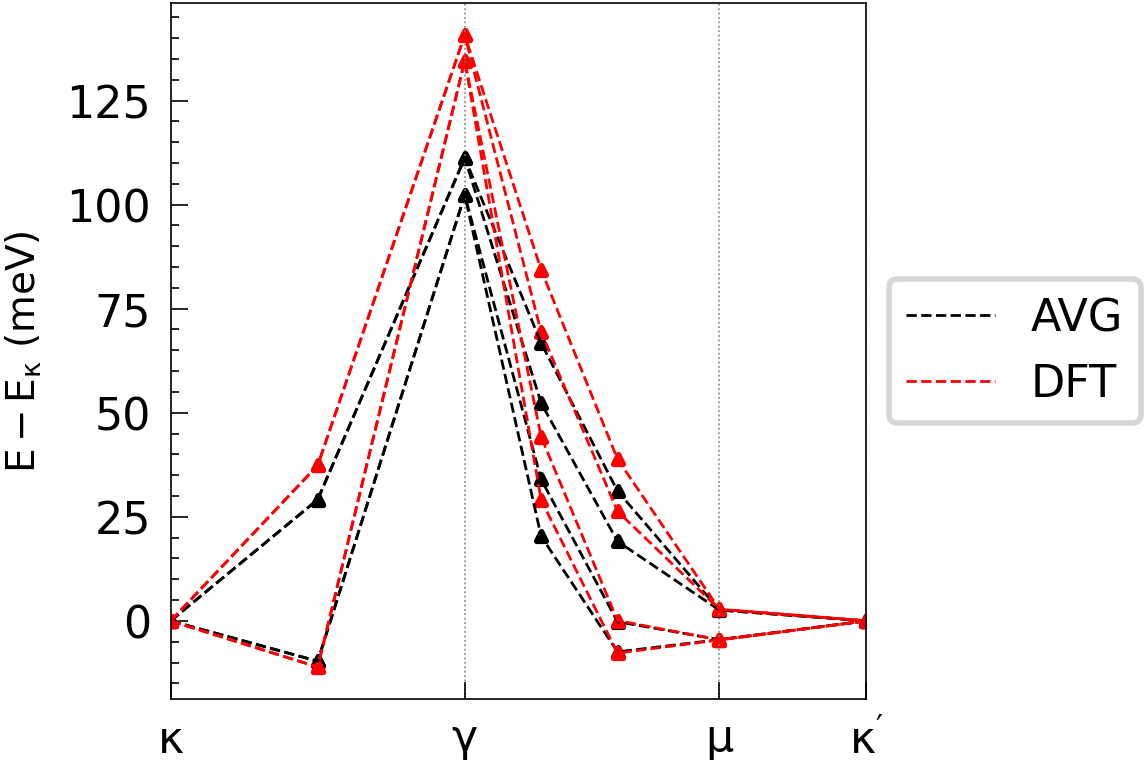}
  \caption{Band structure of ``average" subtraction and DFT subtraction. Zero energy corresponds to moir\'e $\kappa$.}
  \label{fig:direct-comparison}
\end{figure*}

The key qualitative mechanism behind the discrepancy between $\nu=+2$ and $\nu=-2$ is a momentum-dependent single-particle renormalization introduced by the subtraction procedure in the effective interacting Hamiltonian.
In previous studies, it is common to adopt a particle-hole symmetric ``average'' reference density within the active manifold, which suppresses particle-hole asymmetry in the resulting single-particle part and simplifies the mean-field problem.
Within a KS-DFT-based embedding at a fixed reference filling (here charge neutrality), however, such an average reference is far from the projected KS density in the active subspace, since the conduction flat bands are empty in the KS reference state.
In this setting, using a reference density $P_{\mathrm{ref}}$ consistent with the KS-DFT filling is the natural choice for defining $H_{\mathrm{sub}}$.

When the subtraction term is constructed from a realistic reference density $P_{\mathrm{ref}}$ (as opposed to a particle-hole symmetric average reference), the associated Hartree contribution is strongly momentum dependent.
Physically, the projected charge-neutral flat-band density is spatially nonuniform within the moir{\'e} unit cell, and the corresponding Hartree potential therefore produces nonuniform shifts in $\vk$.
This comparison is shown in \cref{fig:direct-comparison}: relative to the average subtraction, the DFT subtraction raises all four active bands near the moir{\'e} $\gamma$ point by about $20$ meV, while leaving the energies near moir{\'e} $\kappa$ much less affected.
Consequently, the relevant indirect gap on the hole-doped side is strongly reduced and can close, yielding a semimetal at $\nu=-2$ even when the underlying KS-DFT band structure is nearly particle-hole symmetric.
Equivalently, the subtraction shifts the energy at moir{\'e} $\gamma$ upward relative to the conduction-band minimum near moir{\'e} $\kappa$, producing an indirect overlap and band crossings at the Fermi level.

The same subtraction-induced renormalization does not drive an analogous band inversion on the electron-doped side in our calculations, so $\nu=+2$ remains insulating and the KIVC solution remains energetically stable.
While the quantitative magnitude of the shift depends on the details of the subtraction scheme, \cref{fig:direct-comparison} shows that the hole-side change is of precisely the scale needed to collapse the indirect gap, and the qualitative tendency toward a hole-side gap reduction is robust as long as the subtraction incorporates the realistic spatial structure of the active-space reference density $P_{\mathrm{ref}}$.

\section{Conclusion}
In this work, we presented an \textit{ab initio} quantum-embedding framework for deriving and solving interacting models of moir{\'e} materials, and demonstrated its application to MATBG at even-integer fillings.
Starting from KS-DFT, we compute screened interactions within cRPA and construct an embedding Hamiltonian in which the double-counting subtraction is defined by a reference density consistent with the KS-DFT filling.
We developed an automated SCDM-based gauge-fixing procedure, which produces a symmetry-adapted localized basis suited for initializing and interpreting solutions from the embedding Hamiltonian.

We recover insulating KIVC solutions at $\nu=0$ and $\nu=+2$, with small CCSD correlation energies beyond HF, so these fillings mainly serve as benchmarks for the embedding workflow against established KIVC physics.
In contrast, at $\nu=-2$ we find a fragile semimetallic solution with a weak $\sqrt{3}\times\sqrt{3}$ Kekul\'e modulation and enhanced intervalley-scattering peaks in the FT-LDOS, consistent with the phenomenology reported in ultra-low-strain STM studies.
Within our purely electronic model, the appearance of enhanced intervalley peaks at $\nu=-2$ can be understood without requiring a fully gapped KIVC insulator, since the symmetry-based arguments enforcing a vanishing Kekul\'e signature apply most directly to gapped phases.

We trace the particle-hole asymmetry between $\nu=\pm 2$ primarily to the subtraction procedure.
When $H_{\mathrm{sub}}$ is constructed from a reference density consistent with the KS-DFT filling, the resulting momentum-dependent Hartree renormalization shifts the valence manifold upward near the moir{\'e} $\gamma$ point and strongly reduces the hole-side indirect gap, driving semimetallic behavior at $\nu=-2$.
These results highlight that, at meV energy scales, subtraction choices and reference densities are not merely technical details and can qualitatively affect the predicted low-energy spectrum. 
The four-band downfolding is least controlled approximation in the pipeline. On the hole-doped side, the separation between the active manifold and the nearest remote bands in the underlying DFT spectrum is around $20$ meV, which is comparable to the subtraction-induced upward shift near moir{\'e} $\gamma$. We therefore cannot exclude the possibility that hybridization with remote-bands can modify, or even remove, the $\nu=-2$ semimetal found in the present treatment. Establishing the robustness of this hole-side semimetal will require an enlarged active-space calculation with controlled gauge fixing and subtraction in the presence of remote bands. One promising direction along this line is the heavy fermion model~\cite{song2022magic,merino2025interplay}. Other future directions include the investigation of strained geometries, a more detailed comparison with recent parameterization of continuum models~\cite{kang2023pseudomagnetic,carr2019exact,QuinnKongLuskinEtAl2025}, and ab initio treatment of electron-phonon interactions~\cite{kwan2024electron}.

\subsection*{Acknowledgments}
This work was supported by the Simons Targeted Grants in Mathematics and Physical Sciences on Moir\'e Materials Magic (R.K., K.D.S., L.L.), by the U.S. Department of Energy, Office of Science, Office of Advanced Scientific Computing Research's Applied Mathematics Competitive Portfolios program under Contract No. AC02-05CH11231, and by the Simons Investigator in Mathematics award through Grant No. 825053 (L.L.) and by the Theory of Materials Program (W.K., S.G.L.) funded by the U.S. Department of Energy, Office of Basic Energy Sciences, under Contract No. DE-AC02-05CH11231 at the Lawrence Berkeley National Laboratory. 
Computational resources were provided by the National Energy Research Scientific Computing Center (NERSC), which is supported by the Office of Science of the U.S. Department of Energy under Contract No. DEAC02-05CH11231, Stampede3 at the Texas Advanced Computing Center (TACC), through Advanced Cyber infrastructure Coordination Ecosystem: Services \& Support (ACCESS), which is supported by National Science Foundation under Grant No. ACI1053575, and Frontera at TACC, which is supported by the National Science Foundation under Grant No. OAC1818253. 
We thank Dumitru C{\u{a}}lug{\u{a}}ru, Mitchell Luskin, Allan MacDonald, Mit Naik, Siddharth Parameswaran, Daniel Parker, Tomohiro Soejima, Alexander Watson, Michael Zaletel for valuable discussions at various stages of this work.

\bibliographystyle{apsrev4-1}
\bibliography{bibliography_TBG}

\newpage
\appendix
\renewcommand\thesubsection{\thesection.\arabic{subsection}}
\crefalias{section}{appendix}
\crefalias{subsection}{appendix}

\begin{center}
\textbf{\Large Supplementary Material for ``Ab Initio Quantum Embedding of Twisted Bilayer Graphene''}
\end{center}

The appendices are organized as follows: \cref{sec:dft-calc-details} summarizes the first-principles simulations and the cRPA screening; \cref{app:gauge-fixing-impl} gives implementation details of the adaptive gauge fixing; \cref{app:init} collects the structured initializations and symmetry diagnostics; \cref{app:dos} defines the LDOS/FT-LDOS; and \cref{sec:randomized-tests} reports additional Hartree-Fock results using unbiased randomized initializations.

\section{First-Principles Simulation and Screening Details}
\label{sec:dft-calc-details}

\subsection{Atomic Relaxation and Density Functional Theory Calculation}
We constructed a rigidly twisted bilayer graphene by rotating the top layer by $1.08^\circ$ around the hexagonal center of the lattice, aligning it with the bottom layer so that their hexagonal centers coincide.
We used 2.46~\AA{} for the graphene lattice constant. Atomic structural relaxation was performed using the LAMMPS package~\cite{thompson2022lammps}.
The REBO force field~\cite{brenner2002second} was used to model the intralayer interatomic potential, while the Kolmogorov Crespi (KC) potential~\cite{kolmogorov2005registry}  was employed for the interlayer interactions.
Structural optimization was carried out with a force tolerance of $10^{-4}~\mathrm{eV}/\text{\AA}$.

Using the relaxed atomic structure, we performed first-principles density functional theory (DFT) calculations with the SIESTA code~\cite{soler2002siesta}.
We adopted the local density approximation (LDA) with Perdew-Zunger parametrization~\cite{perdew1981self} and utilized optimized norm-conserving Vanderbilt (ONCV) pseudopotentials~\cite{hamann2013optimized}.
To capture the interlayer hybridization between the two graphene layers, we employed an expanded basis set that included diffuse 3$s$ and 3$p$ orbitals of carbon atoms.
We optimized the basis set parameters in the SIESTA calculations to ensure that the band structure of AA-stacked bilayer graphene matched that obtained from plane-wave DFT calculations performed with the Quantum ESPRESSO package~\cite{giannozzi2009quantum}.

We converged the self-consistent charge density using a 3$\times$3$\times$1 Monkhorst-Pack $k$-grid and conducted a non-self-consistent band calculation to generate the Kohn-Sham (KS) eigenvalues and eigenstates on a 6$\times$6$\times$1 k-grid.
We observed that with a $\gamma$-sampled charge density, the overall band structure exhibited more dispersive hole bands compared to that obtained with finer k-grid sampling.
We attributed this difference to the singularity of flat bands near the $\gamma$ point in the Brillouin zone.
After completing the DFT calculations, we obtained the KS eigenstates in the plane-wave basis using the \texttt{siesta2bgw} code included in the BerkeleyGW package~\cite{deslippe2012berkeleygw}.

\subsection{Screened Coulomb Interactions}
\label{sec:scre-coul-inter}
The screened Coulomb interaction is a key component of the \textit{ab initio} interacting Hamiltonian. In contrast to continuum models, which often assumes a uniform dielectric constant, our approach derives the interaction from first principles, capturing the material's intrinsic screening. To model the intrinsic screening arising from the higher-energy inactive subspace, we adopt the constrained random phase approximation (cRPA)~\cite{Aryasetiawan2004crpa}. In this approach, the interactions between orbitals in the active space are screened by the static RPA dielectric function, but with screening contributions from virtual transitions within the active subspace explicitly subtracted to avoid double-counting. Consequently, the inverse dielectric matrix at the cRPA level $\varepsilon_{\rm cRPA}^{-1}$ is defined as:
\begin{align}
  \varepsilon_{\rm cRPA}^{-1}(\bvec{q};\bvec{G},\bvec{G}')=\big(\delta_{\bvec{G},\bvec{G}'}-V_{\mathrm{slab}}(\bvec{q}+\bvec{G})\,\chi_{R}(\bvec{q};\bvec{G},\bvec{G}')\big)^{-1},
\end{align}
where the ``rest'' polarizability matrix $\chi_{R}$ is given by:
\begin{align}
   \chi_{R} \equiv \chi_{0} - \chi_{A}.
\end{align}
Here, $\chi_{0}$ represents the total irreducible polarizability of the system, including virtual excitations across the full energy spectrum, while $\chi_{A}$ accounts for the polarizability solely within the flat-band subspace. By removing the metallic screening channels of the flat bands, $\chi_{R}$ is dominated by transitions across a finite energy gap (on the order of tens of meV); thus, $\varepsilon_{\rm cRPA}$ effectively describes the screening environment of a small-gap semiconductor~\cite{Aryasetiawan2004crpa}.

Using the Adler-Wiser expression, the irreducible polarizability matrix of the system can be expressed in terms of the KS eigenvalues and eigenstates:
\begin{align}
  \chi_{0}(\bvec{q};\bvec{G},\bvec{G}') = \frac{1}{N_{\bvec{k}}}\sum_{\bvec{k}}\sum_{\substack{v\in{\text{occ}} \\ c\in{\text{emp}}}}{\frac{\varrho_{v (\bvec{k}+\bvec{q}), c \bvec{k}}^*\left(\bvec{G}\right) \varrho_{v (\bvec{k}+\bvec{q}), c \bvec{k}}\left(\bvec{G}'\right)}{\epsilon_{v (\bvec{k}+\bvec{q})}- \epsilon_{c \bvec{k}}}},
\end{align}
where we recall the pair product \(\varrho_{m \bvec{k}, n \bvec{k}'}(\bvec{G})\) (see~\cref{eq:pair-product-def}),
and ``occ'' and ``emp'' represent the occupied and unoccupied states of the system. For $\chi_{A}$, the band summation is performed only within the flat-band active subspace. The number of unoccupied bands required to converge $\chi_{0}$ is on the order of the number of plane-wave basis functions used in the simulation, making brute-force summation infeasible for large-scale calculations. Inspired by the approach in~\cite{Xuan2019XAFGW}, we apply the following decomposition:

\begin{align}
    \chi_{0} \approx \chi^{H}_{\text{wid}} + \sum_{\text{sub}=\text{top,bot}}\left(\chi^{\text{sub}}_{0} - \chi^{\text{sub}}_{\text{wid}}\right),
\end{align}
where, $\chi^{\text{sub}}_{0}$ is the polarizability matrix of the subsystem (in our case, the pristine top and bottom layer of graphene), and $\chi^{H}_{\text{wid}}$, $\chi^{\text{sub}}_{\text{wid}}$ are the polarizability matrices of heterostructure and individual subsystems, respectively, but only account for the virtual transitions within an energy window:

\begin{align}
  \chi_{\text{wid}}(\bvec{q};\bvec{G},\bvec{G}') = \frac{1}{N_{\bvec{k}}}\sum_{\bvec{k}}\sum_{\substack{v,c \\ \epsilon_{v (\bvec{k}+\bvec{q})},\,\epsilon_{c \bvec{k}}\in[\epsilon_{\min}, \epsilon_{\max}]}}{\frac{\varrho_{v (\bvec{k}+\bvec{q}), c \bvec{k}}^*\left(\bvec{G}\right) \varrho_{v (\bvec{k}+\bvec{q}), c \bvec{k}}\left(\bvec{G}'\right)}{\epsilon_{v (\bvec{k}+\bvec{q})} - \epsilon_{c \bvec{k}}}},
\end{align}
where we select the energy window $[\epsilon_{\min}, \epsilon_{\max}]$ around the Fermi level. 

Therefore, the final expression for the screened Coulomb interaction incorporated into our model is given by:
\begin{align}
  W(\vq;\vG,\vG')=\varepsilon^{-1}_{\rm cRPA}(\vq;\vG,\vG') V_{\mathrm{slab}}(\vq+\vG')
\end{align}
For computational efficiency, we use only the diagonal part of the dielectric screening matrix $\varepsilon^{-1}_{\rm cRPA}(\vq;\vG,\vG')$.

As our calculations employ a slab geometry with periodic boundary conditions, we eliminate spurious interactions between periodic images along the out-of-plane z-direction by substituting the standard 3D Coulomb interaction with the truncated potential $V_{\text{slab}}$~\cite{ismail2006truncation}. The truncated Coulomb interaction correctly models the 2D nature of the slab and can be written as:
\begin{align}
    V_{\text{slab}}(\vq+\vG') = \frac{4\pi}{|\vq+\vG'|^2}\left(1-e^{-|(\vq+\vG')_{xy}|z_c}\cos(|(\vq+\vG')_{z}|z_c)\right)
\end{align}
where $z_c=L_z/2$ is the cut-off length in the $z$-direction, and subscripts $xy$ and $z$ denote the in-plane and out-of-plane components of the vector, respectively.

\subsection{Screened Coulomb and Exchange Matrix Elements}
\label{subsec:form-coul-exch}
Given a (translation-invariant) Hartree-Fock Slater determinant \(\ket{\Phi}\), we define the 1-RDM \(P(\bvec{k})\) by
\([P(\bvec{k})]_{nm} := \braket{\Phi | \hat{f}_{m \bvec{k}}^{\dagger} \hat{f}_{n \bvec{k}} | \Phi }\).
Using Wick's theorem together with translation invariance, we have
\begin{equation}
  \begin{split}
    & \braket{\Phi | \hat{f}_{m \bvec{k}}^{\dagger} \hat{f}_{m' \bvec{k}'}^{\dagger} \hat{f}_{n' (\bvec{k}' - \bvec{q})} \hat{f}_{n (\bvec{k} + \bvec{q})} | \Phi } \\[1ex]
    &\quad = \delta_{\bvec{q},\bvec{0}}\,[P(\bvec{k})]_{n m}\,[P(\bvec{k}')]_{n' m'}
    - \delta_{\bvec{k}',\bvec{k}+\bvec{q}}\,[P(\bvec{k})]_{n' m}\,[P(\bvec{k}+\bvec{q})]_{n m'}.
  \end{split}
\end{equation}
Therefore, in our \textit{ab initio} framework, the Hartree and exchange matrices can be written in terms of the screened Coulomb interaction $W(\vq;\vG,\vG')$ as
\begin{equation}
  \begin{split}
    & [J[P_{\mathrm{ref}}](\bvec{k})]_{mn} \\
    & = \frac{1}{N_{\bvec{k}} |\Omega|} \sum_{\bvec{G}, \bvec{G}'}^{} W(\bvec{0}; \bvec{G}, \bvec{G}') \left( \sum_{\bvec{k}'}^{} \Tr{\Big( \varrho_{\bvec{k}', \bvec{k}'}(\bvec{G})^{\dagger} P_{\mathrm{ref}}(\bvec{k}') \Big)} \right) \varrho_{m \bvec{k}, n \bvec{k}}(\bvec{G}') 
  \end{split}
\end{equation}
and
\begin{equation}
  \begin{split}
    & [K[P_{\mathrm{ref}}](\bvec{k})]_{mn} \\
    & = -\frac{1}{N_{\bvec{k}} |\Omega|} \sum_{\bvec{q}, \bvec{G}, \bvec{G}'}^{} W(\bvec{q}; \bvec{G}, \bvec{G}') \sum_{m' n'}^{} \varrho_{m \bvec{k}, n' (\bvec{k} + \bvec{q})}(\bvec{G})\,[P_{\mathrm{ref}}(\bvec{k} + \bvec{q})]_{n', m'}\,\varrho_{m' (\bvec{k} + \bvec{q}), n \bvec{k}}(-\bvec{G}') .
  \end{split}
\end{equation}

\section{Adaptive Gauge Fixing: Implementation Details}
\label{app:gauge-fixing-impl}
In this section, we describe the numerical implementation of the adaptive gauge-fixing procedure based on selected columns of the density matrix (SCDM). This procedure constructs a localized Wannier-like basis from the underlying \textit{ab initio} Bloch states. We also describe the alignment procedure used to enforce gauge continuity across the Brillouin zone.

For a given $\vk$-point, we represent the set of $N_b$ Bloch bands on a given grid of size $N_g = N_x \times N_y \times N_z$ which yields a matrix $\Psi_\vk \in \mathbb{C}^{N_g \times N_b}$. The goal is to find a unitary transformation $U_\vk$ such that the transformed orbitals are localized along the dominant degree of freedom of the system.
We briefly review the steps in SCDM for finding such a gauge transformation below (see also~\cref{alg:scdm}).

First, we perform a QR factorization with column pivoting (QRCP) on the Hermitian conjugate of the wavefunction matrix,
\begin{align}
  \Psi_\vk^\dagger \Pi = QR,
\end{align}
where $\Pi$ is the permutation matrix, $Q$ is unitary, and $R$ is upper triangular with diagonal elements sorted in descending magnitude.

Second, we identify the ``significant" columns by selecting the first $N_b$ indices from the permutation matrix $\Pi$. Let $\mathcal{C}$ denote the set of selected indices. Physically, these correspond to the grid points where the electron density is most representative.

Then, we construct a set of localized but non-orthogonal orbitals by projecting the full density matrix $P_\vk = \Psi_\vk \Psi_\vk^\dagger$ onto the selected columns $\mathcal{C}$,
\begin{align}
    \tilde{\Phi}_\vk = P_\vk[:,\mathcal{C}] = \Psi_\vk (\Psi_\vk^\dagger)[:,\mathcal{C}]
\end{align}
where $\tilde{\Phi}_\vk$ is the localized orbital matrix.

For the last step, we apply L\"owdin orthogonalization to enforce orthonormality
\begin{align}
    \Phi_\vk = \tilde{\Phi}_\vk S_\vk^{-\frac{1}{2}}
\end{align}
where $S_\vk = \tilde{\Phi}_\vk^\dagger \tilde{\Phi}_\vk$ is the overlap matrix of the non-orthogonal orbitals.

The SCDM procedure described above is performed independently at each $\vk$-point. While SCDM successfully disentangles the subspace (e.g., separating valleys) locally, the orbital ordering can be inconsistent across the Brillouin zone. For example, the first column of $\Psi_\vk$ might correspond to the $\vK$ valley at one $\vk$-point but to the $\vK'$ valley at another.

To ensure smooth gauge continuity, we enforce consistency relative to a reference point (typically $\vk_{\mathrm{ref}} = \gamma$ or the first point in the path). At the initial $\vk$-point, we verify the physical character of the localized orbitals (e.g., by checking their spatial support) and sort along labels (e.g., indices $1 \dots N_b/2$ for valley $\vK$ and $N_b/2+1 \dots N_b$ for valley $\vK'$ in MATBG).

For a subsequent point $\vk_{i}$, we compute the overlap magnitude matrix $O$ between the derived orbitals $\Phi_{\vk_i}$ and the reference orbitals from the previous step $\Phi_{\vk_{\text{ref}}}$
\begin{align}
O_{mn} = \left| \langle \phi_{m \vk_i} | \phi_{n \vk_{\text{ref}}} \rangle \right|
\end{align}
We then permute the columns of $\Phi_{\vk_i}$ to maximize the diagonal elements of $O$. This ensures that the orbital character corresponding to the degree of freedom (e.g., valley and sublattice for MATBG) remains continuous along the $\vk$-path, enabling the accurate analysis of symmetry-breaking ansatzes.

\begin{algorithm}[H]
  \caption{Valley and Sublattice Localization using SCDM}\label{alg:localization}
  \label{alg:valley-sublattice-localization}
\begin{algorithmic}[1]
  \Require{Kohn-Sham orbitals \(\Psi_{\bvec{k}} \in \mathbb{C}^{N_{g} \times 4}\)}
  \Ensure{Updated orbitals \(\{ \psi_{(\sigma, \tau), \bvec{k}} \in \mathbb{C}^{N_{g}} : \sigma \in \{ A, B \}, \tau \in \{ \vK, \vK' \} \}\) where \(\psi_{\sigma, \tau}\) is concentrated on sublattice \(\sigma\) and valley \(\tau\).}
  \State{\(\{ \tilde{\Psi}_{\bvec{k}} \}_{\bvec{k}}\) \(\leftarrow\) \(\mathcal{F}^{-1}_{\Gamma}\Big(\mathtt{scdm}\Big(\{ \mathcal{F}_{\Gamma} (\Psi_{\bvec{k}}) \}_{\bvec{k}}\Big)\Big)\)} \Comment{\(\mathcal{F}_{\Gamma}\) is the Fourier transform between a reciprocal-space representation and the real-space grid}
  \State{Divide \(\tilde{\Psi}_{\bvec{k}}\) into two sets \(\Psi_{\vK, \bvec{k}}\) and \(\Psi_{\vK', \bvec{k}}\)} \Comment{\(\Psi_{\tau, \bvec{k}} \in \mathbb{C}^{N_{g} \times 2}\) localized on valley \(\tau\)}
  \For{\(\tau \in \{ K, K' \}\)}
  \State{\(\{ \tilde{\Psi}_{\tau, \bvec{k}} \}_{\bvec{k}}\) \(\leftarrow\) \(\mathtt{scdm}(\{ \Psi_{\tau, \bvec{k}} \}_{\bvec{k}})\)}
  \State{Divide \(\tilde{\Psi}_{\tau, \bvec{k}}\) into \(\psi_{(A, \tau), \bvec{k}}\) and \(\psi_{(B,\tau), \bvec{k}}\)} \Comment{\(\psi_{(\sigma,\tau), \bvec{k}} \in \mathbb{C}^{N_{g}}\) localized on sublattice \(\sigma\).}
  \EndFor
  \State{\textbf{return} Updated orbitals \((\psi_{(A, \vK), \bvec{k}}, \psi_{(B, \vK), \bvec{k}}, \psi_{(A, \vK'), \bvec{k}}, \psi_{(B, \vK'), \bvec{k}})\)}
\end{algorithmic}
\end{algorithm}

\begin{algorithm}[H]
  \caption{SCDM for \(\bvec{k}\) resolved, isolated bands}
  \label{alg:scdm}
  \begin{algorithmic}[1]
    \Require{Bloch orbitals \(\{ \Psi_{\bvec{k}} \}_{\bvec{k}} \subseteq \mathbb{C}^{N_{g} \times N_{b}}\), a reference \(\bvec{k}\) point \(\bvec{k}_{\mathrm{ref}}\)}
    \Ensure{Spatially localized orbitals \(\{ \Phi_{\bvec{k}} \}_{\bvec{k}} \subseteq \mathbb{C}^{N_{g} \times N_{b}}\)}
    \For{each \(\bvec{k}\)}
    \State{\(Q, \Pi\) \(\leftarrow\) \texttt{rrqr}(\(\Psi_{\bvec{k}}^{\dagger}\))} \Comment{Rank revealing QR on \(\Psi_{\bvec{k}}^{\dagger}\)}
    \State{\(\mathcal{C}_{\bvec{k}}\) \(\leftarrow\) \(\Pi[: N_{b}]\)} \Comment{Select the first \(N_{b}\) pivots from RRQR as columns}
    \State{\(\tilde{\Phi}_{\bvec{k}}\) \(\leftarrow\) \(\Psi_{\bvec{k}}^{\dagger} (\Psi_{\bvec{k}}[:, \mathcal{C}_{\bvec{k}}]\))} \Comment{Compute selected columns of the density matrix}
    \State{\(\Phi_{\bvec{k}}\) \(\leftarrow\) \(\tilde{\Phi}_{\bvec{k}} (\tilde{\Phi}_{\bvec{k}}^{\dagger} \tilde{\Phi}_{\bvec{k}})^{-\frac{1}{2}}\)} \Comment{L{\"o}wdin orthogonalize the selected columns}
    \EndFor
    \For{each \(\bvec{k} \neq \bvec{k}_{\mathrm{ref}}\)}
    \State{\(O\) \(\leftarrow\) \(\Phi_{\bvec{k}}^{\dagger} \Phi_{\bvec{k}_{\mathrm{ref}}}\)}
    \State{\(\Pi\) \(\leftarrow\) \texttt{maximize\_diagonal}(\(O\))} \Comment{\(\Pi\) is the permutation which maximizes the diagonal entries of \(O\)}
    \State{\(\Phi_{\bvec{k}}\) \(\leftarrow\) \(\Phi_{\bvec{k}} \Pi\)} \Comment{Permute the columns of \(\Phi_{\bvec{k}}\)}
  \EndFor
  \end{algorithmic}
\end{algorithm}

\section{Structured initialization and symmetry diagnostics}
\label{app:init}
In the main text we use both structured and randomized initializations to explore symmetry-broken Hartree-Fock solutions.
For completeness, we collect the initialization prescriptions and the symmetry diagnostics used to classify converged states.

\emph{Symmetries and gauge fixing}: The low-energy Hamiltonian obtained from our KS-DFT workflow commutes with $C_{2z}$ and spinless time-reversal $\mathcal{T}$.
Using the adaptive SCDM-based gauge fixing described in \cref{sec:adapt-gauge-fixing}, we fix the gauge so that the DFT orbitals are sublattice polarized and \cref{eq:tr-and-c2-gauge} holds.
This gauge choice enables direct identification of valley and sublattice degrees of freedom in the active manifold.

\emph{Structured initialization for spinless MATBG}: Due to the gauge fixing procedure, we can define Pauli matrices on the sublattice degree of freedom, $\sigma_{x/y/z}$, and the valley degree of freedom, $\tau_{x/y/z}$, acting in the standard way.
For example, $\sigma_x$ exchanges the two sublattices and is defined by $\sigma_x \psi_{(A,\vK),\bvec{k}} = \psi_{(B,\vK),\bvec{k}}$.
Additionally, we set $\sigma_0=\tau_0=I$.

Following Ref.~\cite{bultinck2020ground}, we consider five symmetric candidate states: valley polarized (VP), valley Hall (VH), quantum Hall (QH), Kramers intervalley coherent (KIVC), and time-reversal symmetric intervalley coherent (TIVC).
The corresponding spinless one-body reduced density matrices (1-RDMs) $P_0$ can be written in the valley/sublattice Pauli basis as
\begin{equation}
  \label{eq:initial-states}
  \begin{split}
    &P_{\mathrm{VP}} = \frac{1}{2} ( I + \sigma_{0} \tau_{z}),  \quad P_{\mathrm{VH}} = \frac{1}{2} ( I + \sigma_{z} \tau_{0}), \quad P_{\mathrm{QH}} = \frac{1}{2} (I + \sigma_{z} \tau_{z}), \\[1ex]
    &\hspace{3.5em} P_{\mathrm{KIVC}} = \frac{1}{2} (I + \sigma_{0} \tau_{y}), \quad P_{\mathrm{TIVC}} = \frac{1}{2} ( I + \sigma_{0} \tau_{x}).
  \end{split}
\end{equation}

\emph{Symmetry order parameters from sewing matrices}: Another way of identifying symmetry-broken solutions is through the gauge-invariant symmetry order parameter~\cite{FaulstichStubbsZhuEtAl2023,HouSurWagnerEtAl2025}.
For a given antiunitary symmetry $g$, let $D(g)$ denote its representation matrix on the basis of occupied bands.
The corresponding sewing matrix is defined by
\begin{equation}
  [B_{\bvec{k}}(g)]_{mn} = \braket{u_{m,g\bvec{k}}(\bvec{r}), D(g)\conj{u_{n,\bvec{k}}(\bvec{r})}}.
\end{equation}
When $g$ is a symmetry of the single-particle Hamiltonian (restricted to the band subspace of interest), $B_{\bvec{k}}(g)$ is unitary.
Given $B_{\bvec{k}}(g)$, we define
\begin{equation}
  \mathcal{C}(g) =
    \frac{1}{N_{\bvec{k}}}\sum_{\bvec{k}} \| B_{\bvec{k}}(g) P(g \bvec{k}) B_{\bvec{k}}(g)^{\dagger} - \conj{P(\bvec{k})} \|_{F}.
\end{equation}

For MATBG, we consider three antiunitary symmetry operators: (1) $C_{2z}\mathcal{T}$, (2) $\tau_x\mathcal{T}$, and (3) $\tau_y\mathcal{T}$.
The corresponding symmetry indicators for the candidate initial states in \cref{eq:initial-states} are summarized in \cref{tab:nu0}.
\begin{table}[htbp]
  \centering
  
  \renewcommand{\arraystretch}{1.3}
\begin{tabular*}{.46\textwidth}{@{\extracolsep{\fill}}cccc}
\hline \hline & $\tau_x\mathcal{T}$ & $\tau_y\mathcal{T}$ & $C_{2z}\mathcal{T}$ \\
\hline
$\text{VP}$ & 1 & 1 & 0 \\
$\text{VH}$ & 0 & 0 & 1 \\
$\text{QH}$ & 1 & 1 & 1 \\
$\text{KIVC}$ & 1 & 0 & 1 \\
$\text{TIVC}$ & 0 & 1 & 0 \\
\hline \hline
\end{tabular*}
\caption{Symmetry order parameters of the candidate states. Here $1$ corresponds to symmetry breaking and $0$ corresponds to a symmetry being satisfied.}
\label{tab:nu0}
\end{table}

\emph{Structured initialization for spinful MATBG}: The five candidate states above are defined for spinless systems.
For the spinful Hamiltonian (with negligible spin-orbit coupling), we set the interspin blocks of the 1-RDM to zero, $P_{\uparrow\downarrow}=P_{\downarrow\uparrow}\equiv 0$, and write
\begin{equation}
  P =
  \begin{bmatrix}
    P_{\uparrow\uparrow} & 0 \\
    0 & P_{\downarrow\downarrow}
  \end{bmatrix}.
\end{equation}
Given a spinless 1-RDM $P_0$, when initializing at filling \(\nu\) we make the following choices:
\begin{equation}
  \renewcommand{\arraystretch}{1.3}
  \begin{array}{cccl}
    \nu & P_{\uparrow\uparrow} & P_{\downarrow\downarrow} & \text{Comment} \\
    \hline
    0 & P_{0} & P_{0} & \text{Spin symmetric} \\
    -2 & P_{0} & 0 & \text{Spin polarized} \\
    +2 & P_{0} & I  & \text{Down spin fully occupied}
  \end{array}.
\end{equation}
For example, the KIVC initialization at $\nu=0,-2,+2$ takes the block-diagonal form
\begin{equation}
  \begin{array}{ccc}
    \nu = 0 & \nu = -2 & \nu = +2 \\[1ex]
    \begin{bmatrix}
      P_{\mathrm{KIVC}} & \\
                        & P_{\mathrm{KIVC}}
    \end{bmatrix} &
    \begin{bmatrix}
      P_{\mathrm{KIVC}} & \\
                        & 0
    \end{bmatrix} &
    \begin{bmatrix}
      P_{\mathrm{KIVC}} & \\
                        & I
    \end{bmatrix}
  \end{array}.
\end{equation}
Extensive randomized testing at the Hartree-Fock level indicates that, for the fillings considered here, states of this form typically include the lowest-energy solutions (see \cref{sec:randomized-tests}).

\section{Density of states and local density of states}
\label{app:dos}
In the main text, we visualize the local density of states (LDOS) associated with the Hartree-Fock band eigenstates, as well as its Fourier transform (FT-LDOS).
Given band energies $\epsilon_{n\vk}$ and normalized orbitals $\psi_{n\vk}(\vrr)$, we define the LDOS as
\begin{align}
  \rho(\epsilon,\vrr) := \sum_{n\vk} \delta(\epsilon-\epsilon_{n\vk})\,|\psi_{n\vk}(\vrr)|^2.
\end{align}
In our numerical plots, we replace the delta function by a Lorentzian of width $\eta$,
\begin{align}
  \rho(\epsilon,\vrr) \approx \sum_{n\vk} \frac{1}{\pi}\frac{\eta}{(\epsilon-\epsilon_{n\vk})^2+\eta^2}\,|\psi_{n\vk}(\vrr)|^2.
\end{align}
The density of states (DOS) is obtained by spatially averaging the LDOS over the moir\'e unit cell,
\begin{align}
  \rho(\epsilon) := \frac{1}{|\Omega|}\int_{\Omega} \rho(\epsilon,\vrr)\, d\vrr.
\end{align}
To highlight periodic features in LDOS maps, we compute the Fourier transform of a spatially localized patch of $\rho(\epsilon,\vrr)$; we refer to the resulting intensity map as the FT-LDOS.

\begin{figure}[htbp]
    \centering
    \begin{subfigure}{0.48\textwidth}
        \centering
        \includegraphics[width=0.8\textwidth]{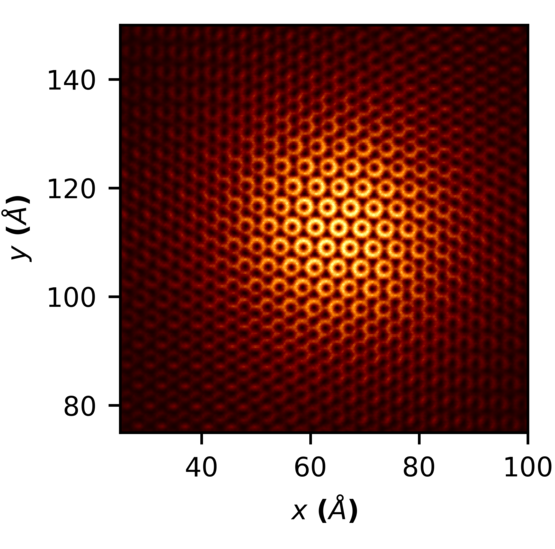}
        \caption{}
        \label{fig:ldos_AA}
    \end{subfigure}
    \begin{subfigure}{0.48\textwidth}
        \centering
        \includegraphics[width=0.87\textwidth]{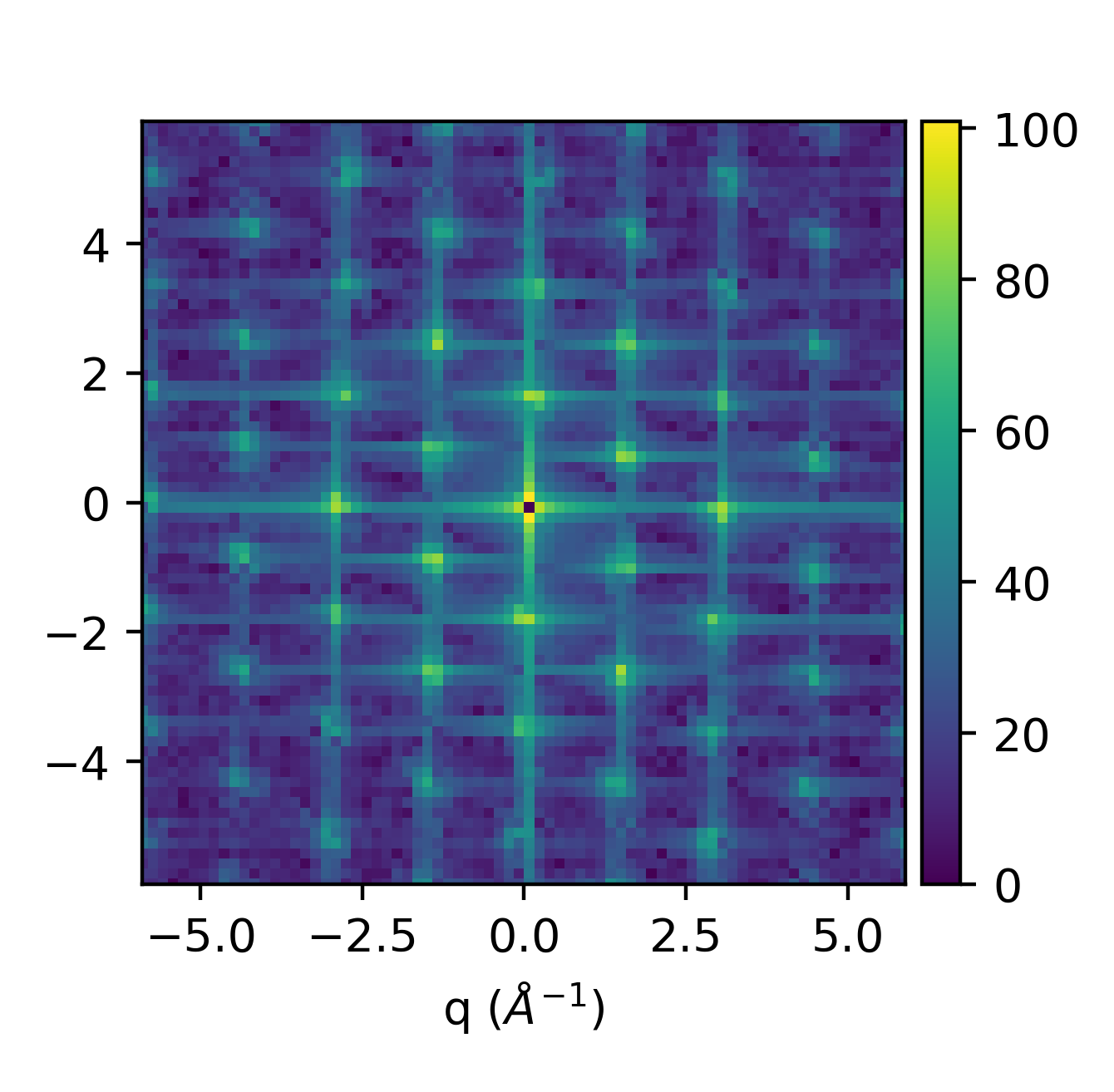}
        \caption{}
        \label{fig:ldos_ft}
    \end{subfigure}
    \caption{(\subref{fig:ldos_AA}) LDOS of the time-reversal symmetric intervalley coherent (TIVC) state in the vicinity of an AA region at charge neutrality. (\subref{fig:ldos_ft}) Fourier transform of a spatially localized patch of the LDOS in (\subref{fig:ldos_AA}). }
    \label{fig:ldos}
\end{figure}
  As an illustrative example, \cref{fig:ldos} shows the LDOS of the time-reversal symmetric intervalley coherent (TIVC) state at charge neutrality, together with the Fourier transform of a spatially localized LDOS patch.
  In \cref{fig:ldos_AA} one observes a Kekul\'e modulation in the LDOS near an AA region, and \cref{fig:ldos_ft} exhibits a corresponding set of additional peaks reflecting this periodicity.

\section{Randomized Tests}
\label{sec:randomized-tests}
\begin{figure}[htbp]
  \centering
  \begin{subfigure}{0.32\textwidth}
    \centering
    \includegraphics[width=\linewidth]{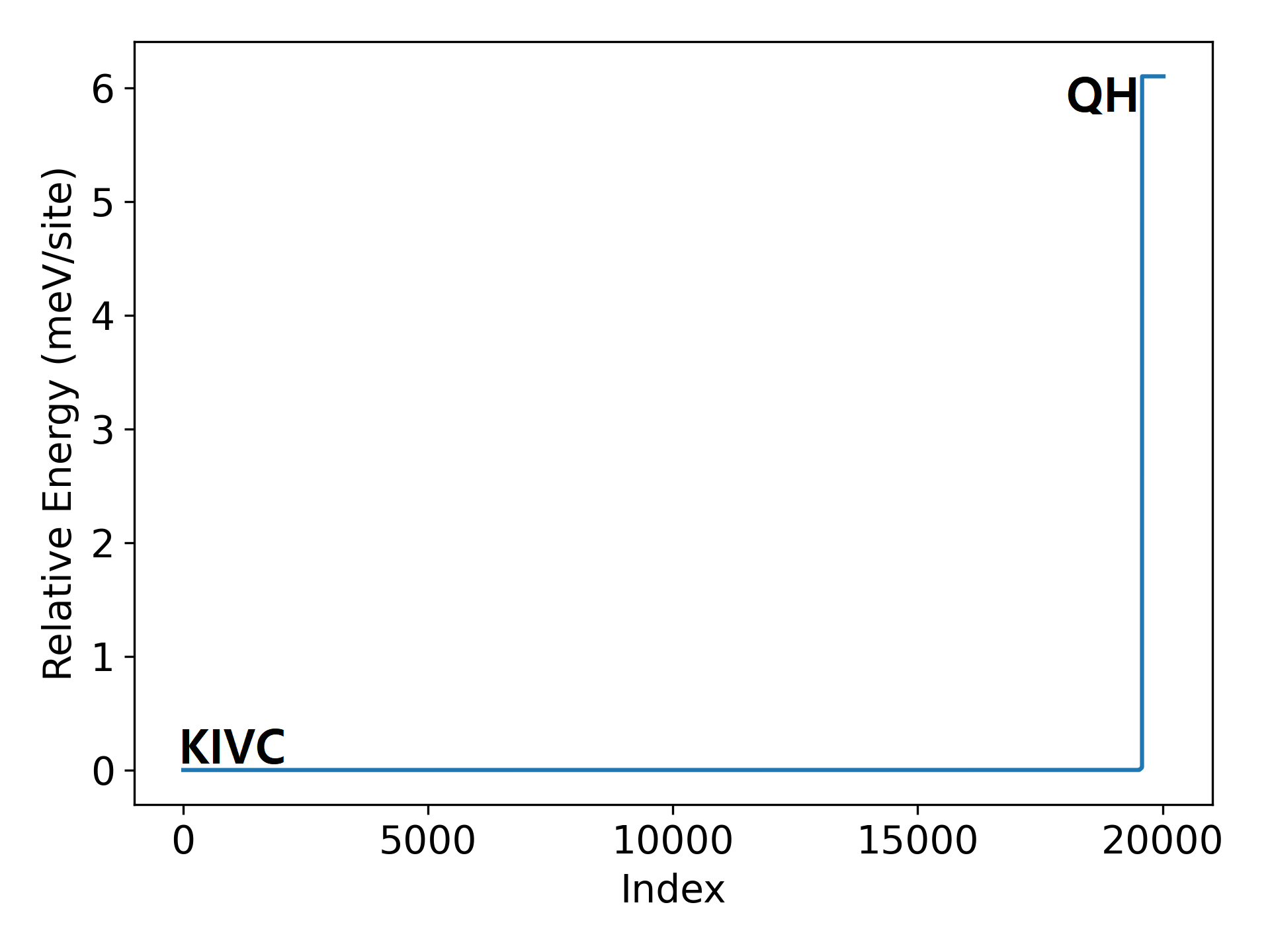}
    \caption{}
    \label{fig:rand_test_cnp}
  \end{subfigure}
  \begin{subfigure}{0.32\textwidth}
    \centering
    \includegraphics[width=\linewidth]{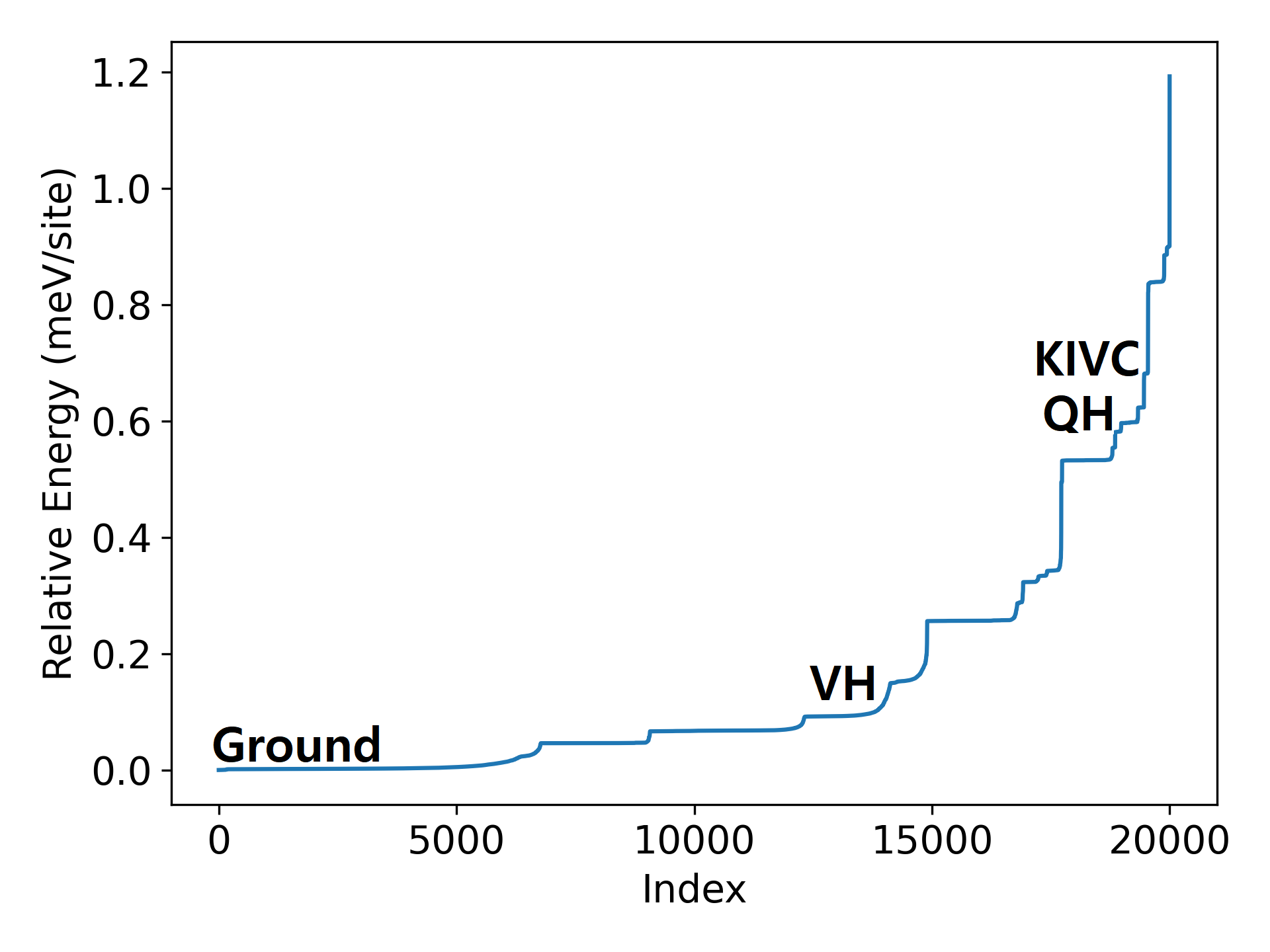}
    \caption{}
    \label{fig:rand_test_doped}
  \end{subfigure}
  \begin{subfigure}{0.32\textwidth}
    \centering
    \includegraphics[width=\linewidth]{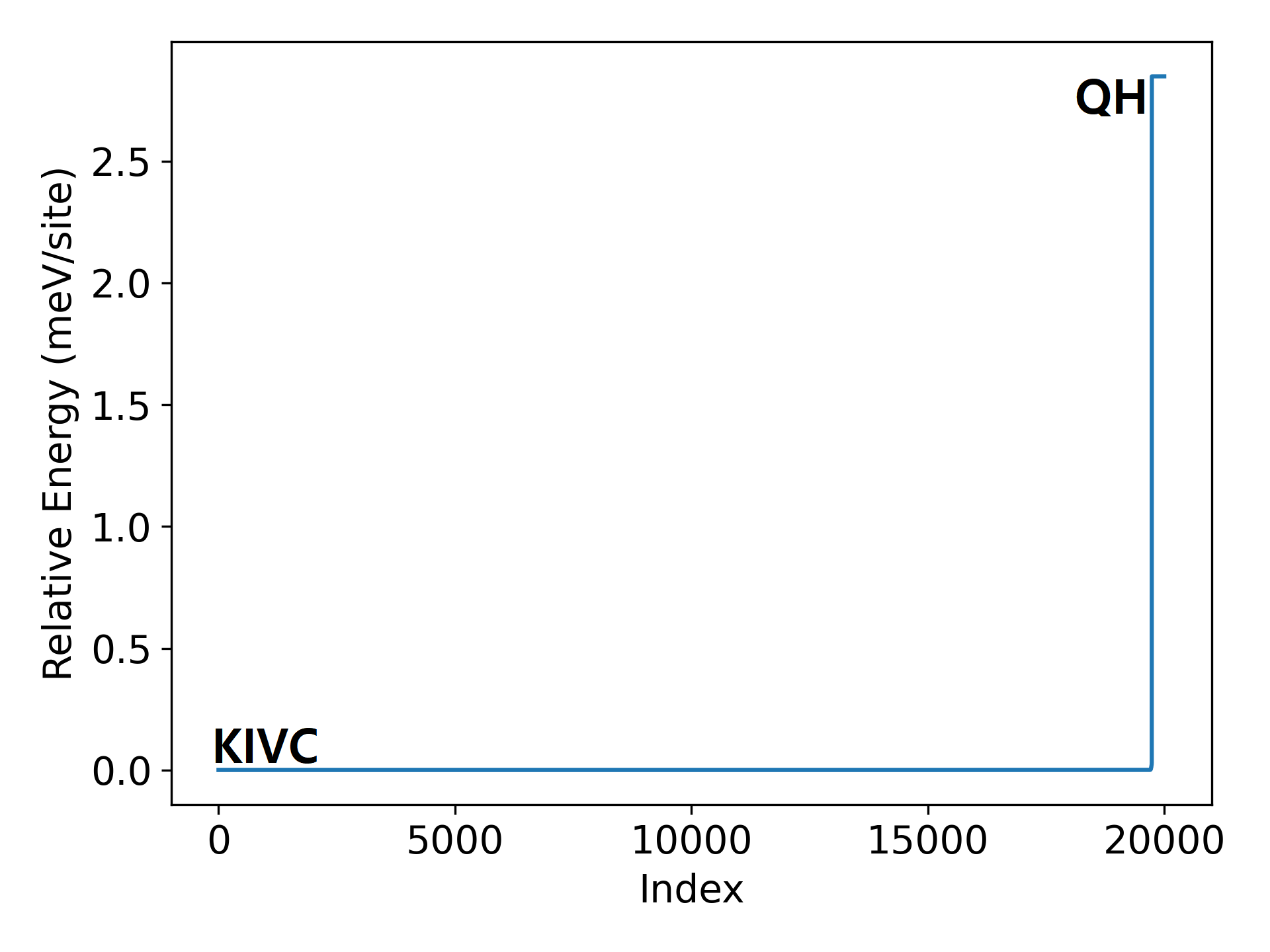}
    \caption{}
    \label{fig:rand_test_particle}
  \end{subfigure}
  \caption{HF energy results with random $P_{\text{init}}$ at (\subref{fig:rand_test_cnp}) $\nu=0$,  (\subref{fig:rand_test_doped}) $\nu=-2$, and  (\subref{fig:rand_test_particle}) $\nu=+2$. The flat plateaus indicate large areas of convergence for specific phases; the lowest-energy plateau corresponds to the lowest-energy solution found in the randomized search. }
  \label{fig:rand_test}
\end{figure}
In the previous sections, we have analyzed the low-energy landscape using a family of five ground states proposed in continuum model studies.
To ensure that our identified low-energy solutions are not merely the lowest-energy states within this limited set of ansatzes, we perform unbiased randomized Hartree-Fock (HF) calculations without prior symmetry assumptions.

For each filling factor, we generate a large set of random initial density matrices ($P_{\text{rand}}$) with the correct number of electrons.
After converging the HF calculation, we sort the final energies to map the solution landscape, as shown in \cref{fig:rand_test}.
We further verify the stability of the resulting HF minima by applying random perturbations of up to 20\% of the density matrix norm.
In all cases, the solutions relax back to their original configurations, confirming that they are robust local minima.

At charge neutrality ($\nu=0$) and particle doping ($\nu=+2$), the randomized search consistently converges to a single dominant minimum (see \cref{fig:rand_test_cnp,fig:rand_test_particle}).
By analyzing the order parameters of the converged states, we confirm that the majority of random initializations converge into the KIVC state. 

On the other hand, the results at hole doping ($\nu=-2$) reveal a more complex landscape.
As shown in \cref{fig:rand_test_doped}, there is a staircase of competing metastable minima appearing in the energy landscape, which aligns with our finding that the $\nu=-2$ phase is a fragile semimetal.
Furthermore, the randomized search finds a lowest-energy solution whose energy and order parameters differ from the standard continuum-model candidates.

\end{document}